\documentclass[useAMS,usenatbib]{mn2e}
\usepackage{graphicx,amsmath,url,natbib}
\usepackage{multirow}

\newcommand{\gtrsim}{\,\raisebox{-0.4ex}{$\stackrel{>}{\scriptstyle\sim}$}\,}
\newcommand{\lesssim}{\,\raisebox{-0.4ex}{$\stackrel{<}{\scriptstyle\sim}$}\,}
\newcommand{\ergcms}{\ensuremath{\mbox{ erg}\mbox{ cm}^{-2}\mbox{ s}^{-1}}}
\newcommand{\E}[1]{\ensuremath{\times10^{#1}}}

\title{The magnetic field strengths of accreting millisecond pulsars}
\author[Mukherjee, Bult, van~der~Klis and Bhattacharya]{%
    Dipanjan Mukherjee$^{1}$\thanks{E-mail:dipanjan.mukherjee@anu.edu.au}, 
    Peter Bult$^2$, 
    Michiel van der Klis$^2$ and 
    Dipankar Bhattacharya$^{3}$ \\
    $^{1}$Research School of Astronomy and Astrophysics, The Australian National University, Canberra, ACT 2611, Australia \\
    $^{2}$Anton Pannekoek Institute, University of Amsterdam, Science Park 904, 1098 XH Amsterdam, The Netherlands \\
    $^{3}$Inter University Centre for Astronomy and Astrophysics, Post Bag 4, Pune 411007, India.
}

\begin{document}

\pagerange{\pageref{firstpage}--\pageref{lastpage}} \pubyear{2015}

\maketitle

\label{firstpage}

\begin{abstract}
In this work we have estimated upper and lower limits to the strength of
the magnetic dipole moment of all 14 accreting millisecond X-ray pulsars observed with
the {\it Rossi X-ray Timing Explorer} ({\it RXTE}). For each source we searched  the
archival {\it RXTE} data for the highest and lowest flux levels with
a significant detection of pulsations. We assume these flux levels to correspond to the
closest and farthest location of the inner edge of the accretion disk at which
 channelled accretion takes place. By estimating the accretion rate from the
observed luminosity at these two flux levels, we place upper and lower limits
on the magnetic dipole moment of the neutron star, using assumptions from standard 
magnetospheric accretion theory. 
Finally, we discuss how our field strength estimates can be further improved
as more information on these pulsars is obtained.
\end{abstract}


\section{Introduction}
Magnetic field strengths of neutron stars span a wide range; from $\gtrsim
10^{14}$~G in magnetars via $\sim 10^{12}$~G in most radio pulsars down to
$\sim 10^8$~G in millisecond pulsars.
The millisecond pulsars are thought to attain their fast spin and low magnetic
field strength through accretion in low-mass X-ray binaries \citep{alpar82, radha82, baile89}. 
The discovery of accreting millisecond X-ray pulsars \citep[AMXPs,][]{wijna98}
and transitional millisecond pulsars \citep{papit13b} has lent strong support 
to this picture. However, a full understanding of the evolutionary scenario 
will only emerge through the comparison of both spin and magnetic field distributions 
of the LMXB neutron stars with those of transitional and millisecond radio pulsars. In this paper we
undertake the task of estimating the magnetic field strengths of the AMXP
population in a consistent approach. 

Magnetic field strengths of neutron stars are estimated in a variety of ways. For
non-accreting radio pulsars the field strengths are inferred from the spin-down rate 
due to electromagnetic torque \citep[e.g.][]{ostri69, beski84, spitk06}. For anomalous 
X-ray pulsars, rough estimates of the field can also be obtained by modelling the
non-thermal X-ray spectra with cyclotron and magnetic Compton scattering
processes in the magnetosphere \citep{guver07,guver08}.

For accreting systems, the available methods for estimating the field strength are less
robust. X-ray pulsars with high surface field strengths ($\gtrsim 10^{12}$~G), may 
show resonant electron cyclotron lines in their X-ray spectra, which
give an estimate of the local field strength in the line formation region \citep{cabal12}. 
However, the location of this region is unclear, resulting in
uncertainties in the measured dipole moment. For lower surface field strengths,
as for AMXPs, cyclotron resonances move out of the X-ray band, leaving only
indirect methods for estimating the field strength. 

In this paper we estimate the magnetic field strength of AMXPs using X-ray
observations obtained with the {\it Rossi X-ray Timing Explorer (RXTE)}.
We assume  that the accretion disk is truncated at an inner radius that depends
on the magnetic field strength and the accretion rate. 
Below the truncation radius the disk is disrupted and matter is forced to move along 
the magnetic field to the magnetic polar caps of the neutron star, creating localised
hotspots responsible for X-ray pulsations \citep[see, e.g.,][]{poutanen06}. The detection 
of X-ray pulsations is taken to imply ongoing  magnetically channelled accretion onto the neutron
star, such that the highest and lowest flux levels with detected pulsations 
identify the range of luminosities (and hence accretion rates) over which  such accretion
occurs. Assuming a disk-magnetic field interaction model, these measurements can
then be used to constrain the surface dipole field strength of the neutron star.

To calculate the dipole moment from the set of flux levels, we consider
the \citet{ghosh78, ghosh79} model of disk-magnetic field interaction,
applied in a manner akin to that adopted by \citet{psalt99}. 
We assume  magnetically channelled to mean that the disk truncation radius is outside the
neutron star surface, but smaller than the co-rotation radius \citep{pring72, illar75}, keeping 
in mind that these radii only set an approximate scale for the system. The accretion
disk may not extend all the way down to the neutron star surface and channelled accretion may
persist for a disk truncated outside the co-rotation radius \citep{sprui93, rappa04, bult15}. 
How this choice of radii affects our magnetic field strength estimates is discussed in 
section~\ref{sec.bdiskinteraction}.

We plan the paper as follows: in Sec.~\ref{sec.method} we review
the theory of accretion-disk/magnetic-field interaction, the details of our field
strength estimation method and X-ray analysis; in Sec.~\ref{sec.results} we describe 
the outburst history of the considered sources and the results of our analysis; in Sec.~\ref{sec.discuss}, 
we discuss the uncertainties in our method and how they affect our results; and finally
in Sec.~\ref{sec.summary} we compare our results with previous estimates of the 
magnetic field strength. Technical details of the timing  and spectral 
analysis are presented in Appendix~\ref{sec.timingdetails} and \ref{sec.fitdetails} 
respectively.

\section{Magnetic field strength estimation method}\label{sec.method}
\subsection{Theory of disk-stellar magnetic field interaction}\label{sec.diskinteraction}
For accretion in a steady state, the inner truncation radius depends on
the balance between magnetic and material stresses. Equating the torque 
from the magnetic stresses and the angular momentum flux \citep{ghosh79,
rappa04}\footnote{%
	Viscous stresses at the truncation point  have
    been ignored},
one finds that
\begin{equation}\label{angular_momentum}
\frac{d(\dot{M} r^2 \Omega _K)}{dr} = B_\phi B_p r^2,
\end{equation}
where $\dot{M}$ is the accretion rate, $r$ the distance from the compact
object, $\Omega _K$ the Keplerian angular velocity at $r$, and $B_p$ and
$B_\phi$ the respective poloidal and toroidal components of the magnetic field.
For simplicity we calculate the torques at the truncation radius considering
the spin axis to be aligned to the magnetic dipole, with the accretion disk
being perpendicular to both axes. The toroidal field component is produced due
to shearing of the poloidal fields. Its strength is an uncertain parameter as
it depends on various poorly understood processes like turbulent diffusion and
magnetic reconnection \citep{wang95}. Another uncertain quantity is the radial 
extent of the accretion disk, $\Delta r$, over which matter couples to the stellar magnetic
field and is channelled away from the disk. By expressing these uncertain quantities 
with the boundary layer
parameter, $\gamma _B = (B_\phi/B_p)(\Delta r/r_t)$, the truncation radius
$r_t$ can be related to the poloidal magnetic field and hence the dipole moment
as
\begin{equation}
B_p = {\gamma _B^{-1/2}} \left(G M \dot{M}^2 \right)^{1/4} r_t^{-5/4}, \label{eq.bpoloidal}
\end{equation}
where $G$ is the gravitational constant and $M$ the neutron star mass. Assuming 
a dipolar magnetic field, $B_p(r) = \mu/r^3$, with $\mu$ the magnetic dipole moment, 
the truncation radius is given as
\begin{equation}
r_t = \gamma _B^{2/7} \left(\frac{\mu ^4}{G M \dot{M}^2}\right)^{1/7}. \label{eq.trunc}
\end{equation}
The truncation radius is related to the classical Alfv\'en radius $r_A$ as $r_t = \gamma _B^{2/7} 2^{1/7} r_A$, where 
\begin{equation}\label{eq.alfven}
\left. \begin{aligned}
r_A &= (2G)^{-1/7} B_s^{4/7}M^{-1/7}\dot{M}^{-2/7}R_s^{12/7} \\
 &= 31 \mbox{ km}\left(\frac{B_s}{10^8 \mbox{ G}}\right)^{4/7}\left(\frac{R_s}{10 \mbox{ km}}\right)^{12/7} \\ 
& \times \left(\frac{\dot{M}}{10^{16} \mbox{ g s}^{-1}}\right)^{-2/7} \left(\frac{M}{1.4 M_\odot}\right)^{-1/7} 
\end{aligned}
\quad \right \}
\end{equation}
Here $B_s=\mu/R_s^3$ is the magnetic field strength at the equator 
and $R_s$ the neutron star radius.  

At higher accretion rates the truncation radius will be closer to the neutron
star surface. In this work we assume the disk to extend all the way down to
the neutron star surface at the highest accretion rate. At the lowest
accretion rate we assume the truncation radius to be at the co-rotation radius. 
Both radii are rough approximations as true behaviour of the accretion
disk truncation radius depends on uncertain aspects like disk/field coupling
and the local magnetic field topology \citep{kulka13,dange12,dange10,roman08}. 
We discuss the effects and limitations of these assumptions in Sec.~\ref{sec.bdiskinteraction}.

By identifying the highest and the lowest accretion rates with
ongoing magnetic channelling (confirmed by the detection of pulsations), we 
estimate the magnetic field as outlined below.

For all sources we adopt the canonical neutron star mass of $M=1.4~M_\odot$ and
radius $R_s=10$ km. The value of $\gamma_B$ is highly uncertain and depends
on where the accretion disk is truncated. 
To be conservative we take $\gamma _B$ to vary between a wide range of $0.01-1$ \citep{psalt99}. 

\begin{enumerate}
\item
{\bf Lower limit on $\boldsymbol{\mu}$:} 
At the highest accretion rates, to observe pulsations, the magnetic field must
be at least high enough to truncate the accretion disk at or above the neutron
star surface. Thus by setting $r_t = R_s$ we obtain the lower limit on the dipole
moment as
\begin{equation}
	\mu _{\rm min} = \gamma_B^{-1/2}(GM)^{1/4}\dot{M}_{\rm max}^{1/2}R_s^{7/4},
\end{equation}
%\begin{eqnarray}
% \mu _{\rm min}  &= & \gamma _B^{-1/2}(GM)^{1/4}\dot{M}_{\rm max}^{1/2}R_s^{7/4}, %  \\
%&=&  1.17 \times 10^{25} \mbox{ G cm}^{-3} \gamma _B^{-1/2} \left(\frac{M}{1.4~M_\odot}\right)^{1/4} \nonumber \\
%&&\times \left(\frac{\dot{M}_{\rm max}}{10^{16} \mbox{ g s}^{-1}}\right)^{1/2} \left(\frac{R_s}{10 \mbox{ km}}\right)^{7/4} \\
%&= &  8.55 \times 10^{24}  \mbox{ G cm}^{-3}  \gamma _B^{-1/2} \left(\frac{M}{1.4~M_\odot}\right)^{-1/4} \nonumber  \\
%&&\times \left(\frac{ \epsilon _{\rm bol} L_{\rm max}}{10^{36} \mbox{ erg s}^{-1}} \right)^{1/2} \left(\frac{R_s}{10 \mbox{ km}}\right)^{9/4} \label{eq.mumin}
%\end{eqnarray}
We assume the mass accretion rate can be estimated from the bolometric luminosity
as $L = GM\dot{M}/R_s$, and estimate $L$ from the observed luminosity in the
X-ray band by applying a bolometric correction factor ($L=\epsilon_{\rm bol}L_X$).
The typical reported values of the bolometric correction factor have a range
of $\epsilon _{\rm bol} \sim 1-2$ \citep{gilfa98, gallo02, campa03, migli06, 
casel08, gallo08}. 
The mass accretion rate then follows as
\begin{align}
	\dot{M} 
		&= 10^{16} \mbox{ g s}^{-1} 
					\left( \frac{ \epsilon _{\rm bol} L_X}{1.87 \times 10^{36}\mbox{ erg s}^{-1}} \right) \nonumber \\
		& \times 	\left( \frac{M}{1.4~M_\odot} \right)^{-1} 
					\left(\frac{R_s}{10 \mbox{ km}}\right).
\end{align}
Expressing the X-ray luminosity in terms of the observed X-ray flux and
the assumed distance  ($L_X = 4 \pi d^2 F$), we obtain the lower limit
on the magnetic moment
\begin{align}
	\mu_{\rm min} &=  9.36 \times 10^{24}  \mbox{ G cm}^{-3}  
		\left( \frac{ \gamma_B }{ 1 } \right)^{-1/2} 
		\left( \frac{ \epsilon_{\rm bol} }{ 1 } \right)^{1/2} 
	\nonumber  \\ &\times	
		\left( \frac{ F_{\rm max}}{10^{-10}~\ergcms} \right)^{1/2} 
		\left( \frac{d}{10 \mbox{ kpc}}\right) 
	\nonumber  \\ &\times	
		\left(\frac{M}{1.4~M_\odot}\right)^{-1/4}
		\left( \frac{R_s}{10 \mbox{ km}}\right)^{9/4} ,
		\label{eq.mumin}
\end{align}
where $F_{\rm max}$ is the highest observed X-ray flux with pulsation and
we adopted the boundary values of $\gamma_B$ and $\epsilon_{\rm bol}$ that
give the lowest magnetic moment. For a detailed discussion of these assumptions we
again refer to sec.~\ref{sec.discuss}.

\item
{\bf Upper limit on $\boldsymbol{\mu}$:} 
We assume magnetic channelling to be centrifugally inhibited if the accretion
disk is outside the co-rotation radius
\begin{equation}
	r_c = \left( \frac{GM}{(2\pi\nu_s)^2} \right)^{1/3},
\end{equation}
%\begin{equation}\label{eq.corot}
%	r_c = 77.93 \mbox{ km} \left(\frac{M}{1.4~M_\odot}\right)^{1/3} \left(\frac{\nu _s}{100 \mbox{ Hz}} \right)^{-2/3},
%\end{equation}
where $\nu_s$ is the spin frequency. We then obtain an upper limit on the
dipole moment by setting the truncation radius at the co-rotation
radius ($r_t = r_c$) for the lowest accretion rate with detected pulsations
\begin{equation}
	\mu _{\rm max} = {\gamma _B^{-1/2}}{(2\pi)^{-7/6}} (GM)^{5/6}\dot{M}_{\rm min}^{1/2}\nu _s^{-7/6}.
\end{equation}
%\begin{eqnarray}
%\!\!\!\mu _{\rm max}  &=&  \frac{\gamma _B^{-1/2}}{(2\pi)^{7/6}} (GM)^{5/6}\dot{M}_{\rm min}^{1/2}\nu _s^{-7/6} \\
% &= &  4.25 \times10^{26} \! \mbox{ G cm}^{-3}  \gamma _B^{-1/2}  \left(\frac{M}{1.4~M_\odot}\right)^{5/6}  \nonumber \\ 
%&& \times \left(\frac{\dot{M}_{\rm min}}{10^{16} \mbox{ g s}^{-1}}\right)^{1/2}  \left(\frac{\nu _s}{100 \mbox{ Hz}}\right)^{-7/6} \\
% &= &  9.83 \times 10^{25} \! \mbox{ G cm}^{-3} \gamma _B^{-1/2} \left(\frac{M}{1.4~M_\odot}\right)^{1/3} \nonumber \\ 
%&& \times \left(\frac{ \epsilon _{\rm bol} L_{\rm min}}{10^{35} \mbox{ erg s}^{-1}} \right)^{1/2} \left(\frac{R_s}{10 \mbox{ km}}\right)^{1/2}  \left(\frac{\nu _s }{100 \mbox{ Hz}} \right)^{-7/6} \label{eq.mumax}
%\end{eqnarray}
By again substituting the expression for the mass accretion rate we obtain
the upper limit on the magnetic dipole moment
\begin{align}
	\mu_{\rm max} &= 1.52 \times 10^{27}~\mbox{ G cm}^{-3} 
		\left( \frac{ \gamma_B }{ 0.01 } \right)^{-1/2}
		\left( \frac{ \epsilon_{\rm bol} }{ 2 } \right)^{1/2}
	\nonumber \\ & \times 
		\left( \frac{ F_{\rm min}}{10^{-11}~\ergcms} \right)^{1/2} 
		\left( \frac{d}{10 \mbox{ kpc}} \right) 
	\nonumber \\ & \times 
		\left( \frac{M}{1.4~M_\odot} \right)^{1/3} 
		\left( \frac{R_s}{10 \mbox{ km}} \right)^{1/2}  
		\left( \frac{\nu_s }{100 \mbox{ Hz}} \right)^{-7/6},
	\label{eq.mumax}
\end{align}
where $F_{\rm min}$ is the lowest observed X-ray flux with pulsation and 
we adopted the boundary values of $\gamma_B$ and $\epsilon_{\rm bol}$ that
maximize the magnetic moment.

%We discuss possible corrections to the above relation due to non-dipolar
%distorted magnetic field in Sec.~\ref{sec.summary}.
\end{enumerate}

%For all sources we consider the canonical neutron star mass of $M=1.4 \, M_\odot$ and
%radius $R_s=10$ km. We further take $\gamma _B$ to vary between $0.01-1$, which is
%a conservative choice that is within the range obtained from numerical 
%simulations \citep[$r_t \sim(0.5-1.2) r_A$,][]{zanni09,roman08}. Specifically,
%we calculate the lower limit using $\gamma_B = 1$ and the upper limit using
%$\gamma_B = 0.01$. Then, using eq.~\ref{eq.mumin} and~\ref{eq.mumax} we can constrain the dipole moment 
%and hence the equatorial surface magnetic field ($B_s$) of AMXPs. 

\subsection{Data analysis}\label{sec.data}
We analysed all AMXP outbursts observed with {\it RXTE}. The
general structure of our analysis is as follows; for each 
{\it RXTE} observation we estimate the Crab normalized X-ray flux and 
then search for the presence of pulsations. We select 
the highest and lowest flux observations with detected pulsations, and from
spectral analysis measure the source flux. Using these flux measurements and
the best distance estimate from literature, we obtain limits on the magnetic
dipole moment from eq. \ref{eq.mumin} and \ref{eq.mumax}. These limits are expressed
in terms of the  magnetic dipole field strength at the equator ($B_p = \mu/r$), assuming
a 10 km radius. The detailed procedure is outlined below.

%To estimate the background contribution we also select an observation near the end of the 
%outburst, when the source itself has presumably returned to quiescence, and again compute 
%the bolometric flux from spectral analysis \citep[e.g.][]{paizi05}. 
%The selected RXTE observation
%IDs (ObsIDs) are reported in appendix~\ref{sec.fitdetails}. All quoted flux 
%estimates are corrected for particle background and dead-time effects. 

%The limits to the magnetic moment are derived by applying eq.~\ref{eq.mumin} and 
%eq.~\ref{eq.mumax} using the measured flux, corrected for the contaminating 
%X-ray background, and source distances from literature. 
%Our final results are presented in Table~\ref{table.fieldvalues}.

\subsubsection{Timing analysis} \label{sec.timing}
We initially estimate the 2--16~keV Crab normalised X-ray flux from 
the 16-second time-resolution Standard-2 data (see e.g. \citealt{straaten03}
for details). To search for pulsations we consider the high time resolution 
GoodXenon or ($122~\mu s$) Event mode data of the same observation, selecting only 
those events in energy channels 5--37 ($\sim2-16$~keV), which usually provides
an optimal signal to noise ratio for the pulsations. The data were 
barycentred using the FTOOLS task {\it faxbary}, which also applies 
the {\it RXTE} fine clock corrections, thus allowing for timing analysis at an 
absolute precision of $\sim4\mu$s \citep{rots04}. We then take each $\sim3000$~s
continuous light curve (as set by the {\it RXTE} orbit), and fold it on
the pulsar ephemeris (see~appendix~\ref{sec.timingdetails}) to construct
a pulse profile. For each profile we measure the amplitude at the fundamental
frequency and that of its second harmonic \citep{hartm08}.

In standard procedures (see, e.g, \citealt{patru13}), a pulsation is usually
said to be significant if the measured amplitude exceeds a detection threshold. 
This threshold is set as the amplitude for which there is only a small
probability $\epsilon$ that one the of observations in an outburst exceeds it by chance. 
For observed amplitudes higher than this threshold we have a high 
confidence $\mathcal{C} = 1-\epsilon$ that pulsations are detected ($\mathcal{C}=99.7\%$). We can
then consider the flux estimates associated with the significant pulse detections
and straightforwardly select the observations of highest and lowest flux. 

This approach is very conservative, as it sets a small joint false-alarm probability
of detection for the entire outburst, in spite of the fact that we can be certain that
pulsation are present in most observations. At the low flux end of the outburst, where
the detection significance decreases with the count rate, this may cause us to miss pulsations.

To overcome this issue we first reduce the number of trials by comparing 
the observed total count rate of an observation with the X-ray background 
as estimated with the FTOOLS task {\it pcabackest} \citep{jahod06}. We then 
set a minimal count-rate threshold and reject all observation for which 
the pulse amplitude cannot be detected above the noise level assuming the 
expected source contribution is 100\% modulated. 
%We note that lowering the assumed pulse fraction will increase the count-rate threshold.

We then select all observations of an outburst that do not have significant 
pulsations according to the procedure described above. If the pulsar emission is indeed 
not present in these observations, then the distribution of measured
amplitudes and phases should correspond to the expected distribution of random noise,
i.e. the phases should be uniform and the squared amplitude should follow a $\chi^2$-distribution
for two degrees of freedom.
We compare the distributions using a KS-test, again with a $99.7\%$ confidence level. If
the data is not randomly distributed we take out the highest flux observation
that has a significant pulse detection at the single trial level and whose phase is consistent
with the expectation from the timing model, and iterate until the sample is consistent with 
being random. The last removed observation is then taken as the lowest observed flux with pulsations.
We note that in practice the sensitivity of this iterative approach is limited by
the small number of observations in the tail of an outburst and only rarely yields a
lower flux pulse detection than through the initial procedure outlined above.

\subsubsection{Spectral analysis}
For the spectral analysis of selected observations we used HEASOFT version 6.12
and the calibration database (CALDB). Spectra were extracted from the Standard-2 data following the
standard procedures outlined in the {\it RXTE}
cookbook\footnote{\url{http://heasarc.nasa.gov/docs/xte/recipes/cook_book.html}}.
The background was again estimated using the FTOOLS task {\it pcabackest}. A dead-time 
correction was applied to the spectra following the prescription in the {\it RXTE}
cookbook\footnote{\url{http://heasarc.gsfc.nasa.gov/docs/xte/recipes/pca_deadtime.html}}.
The spectral fits were done in the 3--20~keV energy range with XSPEC
version 12.7.1. 

From the measured flux we calculate the 3--20~keV X-ray luminosity and
convert to the bolometric luminosity by multiplying with the  
correction factor $\epsilon _{\rm bol}$.  The bolometric correction
factor for a source depends on its spectral state, which in turn varies with
accretion rate. The typical range of the bolometric correction factor is
reported to be $\epsilon _{\rm bol} \sim 1-2$ \citep{gilfa98, gallo02, 
campa03, migli06, casel08, gallo08}. To be conservative we use $\epsilon
_{\rm bol}=1$ when calculating $\mu_{\rm min}$ and $\epsilon _{\rm bol}=2$ when
calculating $\mu_{\rm max}$.

For many of the AMXPs considered here, there is considerable contaminating
background emission in the observed X-ray flux, for instance from Galactic ridge
emission. To estimate the background contamination, we also measured the bolometric
luminosity for an observation in the tail of the outburst, where
pulsations were not present and the light curve has asymptotically levelled
off to a constant value. We assume that in such a state, accretion has ceased
and the observed flux is purely due to background emission.  

The details of the spectral fit parameters for different sources are presented in
Appendix.~\ref{sec.fitdetails}. 
%The results of the analysis and estimated magnetic fields for different sources are presented in Sec.~\ref{sec.results}.

% !TEX root =  amxp.tex

\begin{table*}
\centering
\caption{%
    Flux range with pulsations of the analysed AMXPs in order of ascending spin frequency.
}
\label{table.fluxstates}
\begin{tabular}{| l | l | c | l | l | l | l | l | l |}
\hline
   &      &      & \multicolumn{2}{c|}{Maximum flux} & \multicolumn{2}{c|}{Minimum flux} & \multicolumn{2}{c|}{Background flux} \\ 
No & Name & Spin &  MJD & Flux                       & MJD & Flux                        & MJD & Flux      \\
   &      & (Hz) &      & (\ergcms)                  &     & (\ergcms)                   &     & (\ergcms) \\
\hline
1  & Swift~J1756.9--2508  & 182 & 55026.1 & $6.30 \times 10^{-10}$  & 55032.5 & $1.99 \times 10^{-10}$  & 55037.0 & $4.07 \times 10^{-11}$    \\
2  & XTE~J0929--314	      & 185 & 52403.5 & $4.42 \times 10^{-10}$  & 52442.3 & $6.64 \times 10^{-11}$  & --      &	 --                       \\
3  & XTE~J1807.4--294     & 191 & 52697.6 & $8.19 \times 10^{-10}$  & 52808.7 & $3.51 \times 10^{-10}$  & 52816.8 & $7.25 \times 10^{-11}\,^\text{~}$\\
4  & NGC~6440~X-2         & 206 & 55073.1 & $2.62 \times 10^{-10}$  & 55873.3 & $3.36 \times 10^{-11}$  & 55823.2 & $1.34 \times 10^{-11}$    \\
5  & IGR~J17511--3057     & 245 & 55088.8 & $8.65 \times 10^{-10}$  & 55124.0 & $1.00 \times 10^{-10}$  & 55118.2 & $6.96 \times 10^{-11}\,^\text{~}$\\
6  & XTE~J1814--338       & 314 & 52814.3 & $4.41 \times 10^{-10}$	& 52844.1 & $6.00 \times 10^{-11}$  & 52852.9 & $1.00 \times 10^{-12}$    \\
7  & HETE~J1900.1--2455   & 377 & 53559.5 & $1.15 \times 10^{-9 }$  & 53573.8 & $3.84 \times 10^{-10}$  & --      &  --                       \\
8  & SAX~J1808.4--3658    & 401 & 52563.2 & $1.85 \times 10^{-9 }$  & 50936.8 & $2.82 \times 10^{-11}$  & 50935.1 &	$1.21 \times 10^{-11}\,^\text{~}$\\
9  & IGR~J17498--2921     & 401 & 55789.6 & $1.13 \times 10^{-9 }$  & 55818.3 & $4.44 \times 10^{-10}$  & 55826.4 & $4.23 \times 10^{-10}\,^\text{~}$\\
10 & XTE~J1751--305       & 435 & 52368.7 & $1.50 \times 10^{-9 }$  & 52377.5 & $3.97 \times 10^{-10}$  & 52380.7 & $6.51 \times 10^{-11}$    \\
11 & SAX~J1748.9--2021    & 442 & 55222.5 & $4.13 \times 10^{-9 }$  & 52198.3 & $2.96 \times 10^{-9 }$  & 55254.6 & $1.88 \times 10^{-11}$	  \\
12 & Swift~J1749.4--2807  & 518 & 55300.9 & $5.24 \times 10^{-10}$  & 55306.7 & $2.67 \times 10^{-10}$  & 55307.7 & $2.41 \times 10^{-10}\,^\text{~}$\\
13 & Aql~X-1              & 550 & 50882.0 & $8.74 \times 10^{-9 }$  & 50882.0 & $8.74 \times 10^{-9 }$  & 50939.8 & $1.34 \times 10^{-11}$    \\
14 & IGR~J00291+5934      & 599 & 53342.3 & $9.70 \times 10^{-10}$  & 53352.9 & $1.09 \times 10^{-10}$  & 53359.5 & $5.76 \times 10^{-11}\,^\text{~}$\\
\hline
\end{tabular} \\
\flushleft
%(a) The estimated background has not been used in calculating the magnetic field (see text). \\ \medskip
For each source we give the minumum and maximum flux with pulsations.
We also give the corresponding 3--20~keV background flux used in calculating the 
magnetic dipole moment and the MJD of the analysed observations.
\end{table*}

\begin{table*}
\centering
\caption{%
    The magnetic field strength estimates (equatorial surface field) of all
    considered AMXPs. 
}
\label{table.fieldvalues}
\begin{tabular}{| l l l c c | l l l | l l |}
\hline
   %&      &      &           &            & {Lower limit} & \multicolumn{2}{c|}{Upper limit}   & \multicolumn{2}{c|}{Literature} \\
No & Name & Spin & Distance  & Background & $B_{\rm min}$ &  $B_{\rm max}$ &  $B_{\rm K, max}$ &  $B_{L}$   & ref.  \\
   &      & (Hz) & (kpc)     & corrected  & ($10^8$ G)    & ($10^8$ G)     & ($10^8$ G)        & ($10^8$ G) &       \\
\hline                                          	
{1}  & {Swift~J1756.9--2508} & 182   &   $8$   & y & 0.18  & 24.1   & 7.2   & $0.4 - 9$    & a    \\
{2}  & {XTE~J0929--314}      & 185   &   $6$   & n & 0.12  & 11.5   & 3.4   & $-$          &      \\
{3}  & {XTE~J1807.4--294}    & 191   &  $4.4$  & n & 0.11  & 18.6   & 5.4   & $-$          &      \\
{4}  & {NGC~6440~X-2}        & 206   &  $8.2$  & y & 0.12  &  7.6   & 2.1   & $-$          &      \\
{5}  & {IGR~J17511--3057}    & 245   &   $7$   & n & 0.19  & 11.8   & 3.1   & $-$          &      \\
{6}  & {XTE~J1814--338}      & 314   &   $8$   & y & 0.16  &  7.8   & 1.8   & $\lesssim 5 - 10$ & b     \\
{7}  & {HETE~J1900.1--2455}  & 377   &   $5$   & n & 0.16  & 10.0   & 2.1   & $-$          &      \\
{8}  & {SAX~J1808.4--3658}   & 401   &  $3.5$  & n & 0.14  &  1.77  & 0.36  & $0.7 - 1.5$  & c    \\
{9}  & {IGR~J17498--2921}    & 401   &   $8$   & n & 0.20  & 16.0   & 3.2   & $-$          &      \\
{10} & {XTE~J1751--305}      & 435   &   $7$   & y & 0.25  & 11.0   & 2.1   & $3.3 - 4.7$  & d    \\
{11} & {SAX~J1748.9--2021}   & 442   &  $8.2$  & y & 0.49  & 37.8   & 7.2   & $-$          &      \\
{12} & {Swift~J1749.4--2807} & 518   &  $6.7$  & n & 0.11  &  7.7   & 1.4   & $-$          &      \\
{13} & {Aql~X-1}             & 550   &   $5$   & y & 0.44  & 30.7   & 5.3   & $\lesssim 9$ & e    \\
{14} & {IGR~J00291+5934}     & 599   &   $3$   & n & 0.085 &  1.9   & 0.31  & $1.5 - 2.0$  & f    \\
\hline
\end{tabular}\\
\flushleft
(a) \citet{patru10b}, (b) \citet{watts08, papit07, haske11}, (c) \citet{patru12b}, 
(d) \citet{riggi11b}, (e) \citet{disal03},  (f) \citet{patru10c}. \\ \medskip
The values $B_{\rm min}$ and $B_{\max}$ correspond to field strengths estimated
using eq.~\ref{eq.mumin} and eq.~\ref{eq.mumax}. 
The values $B_{\rm K}$ are upper limits to the field strength computed using
eq.~\ref{eq.newgamma} following the modified expression of truncation radius as
obtained by \citet{kulka13} (see Sec.~\ref{sec.bdiskinteraction}).  
$B_L$ are field strength measurements from literature (see references).  
The background correction column indicates if the background estimate
(Table~\ref{table.fluxstates}) was used when calculating the upper limit on the
magnetic field estimate (see, e.g, Sec.~\ref{sec.res.bkg}).
\end{table*}

\section{Results}\label{sec.results}
In the following sections we present the magnetic field estimates we obtained 
from the timing and spectral analysis of all AMXPs observed with {\it RXTE}. For
each source we briefly describe the outburst history, distance estimates and discuss
specific details of our analysis. All results are summarized in Tables~\ref{table.fluxstates}
and~\ref{table.fieldvalues}.
%and identify the ObsIDs corresponding to the highest and 
%lowest flux for which pulsations were detected (see Table~\ref{table.fluxstates}). 

\subsection{Swift J1756.9--2508}
\begin{figure}
	\centering
	\includegraphics[width=\linewidth] {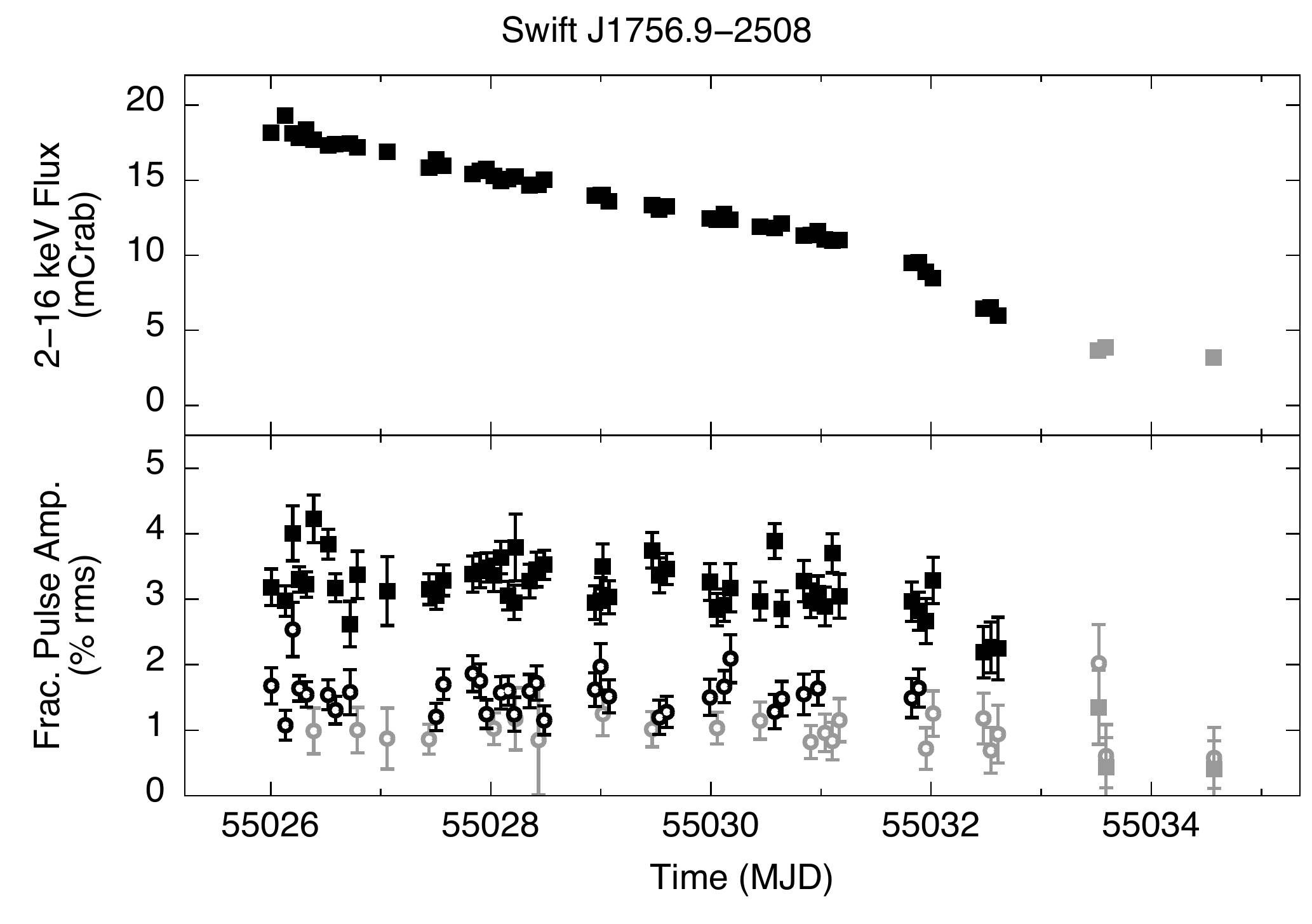}
	\caption{\small  Upper panel: Light curve of the 2009 outburst of Swift J1756.9--2508
        normalised to mCrab. 
        Lower panel: pulse amplitude of the fundamental
        (squares) and the second harmonic (open circles). The observations
        with significant detection (with 99.7\% confidence limit) of pulsation
        are marked with black, observations without a significant detection of
        pulsations are shown in grey.}
	\label{fig.J1756}
\end{figure}

Swift~J1756.9--2508 was first discovered with {\it Swift} in June 2007 
\citep{krimm07a,krimm07b}, and 182~Hz pulsations were found with follow-up
{\it RXTE} observations \citep{markw07b}. The source showed a second outburst 
in July 2009 \citep{patru09c,patru10b}. 

We find both the highest and lowest flux with pulsations to occur during the 
2009 outburst (Fig.~\ref{fig.J1756}), with a detection of pulsation for the
outburst peak luminosity at MJD~55026.1 and lowest flux detection on MJD~55032.5,
just before the light curve decays to the background level.
The background contribution was measured from the last observation on MJD~55037.0. 

The distance to the source is not known, but considering its close proximity 
to the galactic centre \citep{krimm07a}, we assume a distance of $8$~kpc. We then obtain
a magnetic field range of $1.8\E{7}~\mbox{G} < B < 2.4\E{9}$~G.

\subsection{XTE J0929--314}
\begin{figure}
	\centering
	\includegraphics[width=\linewidth]{{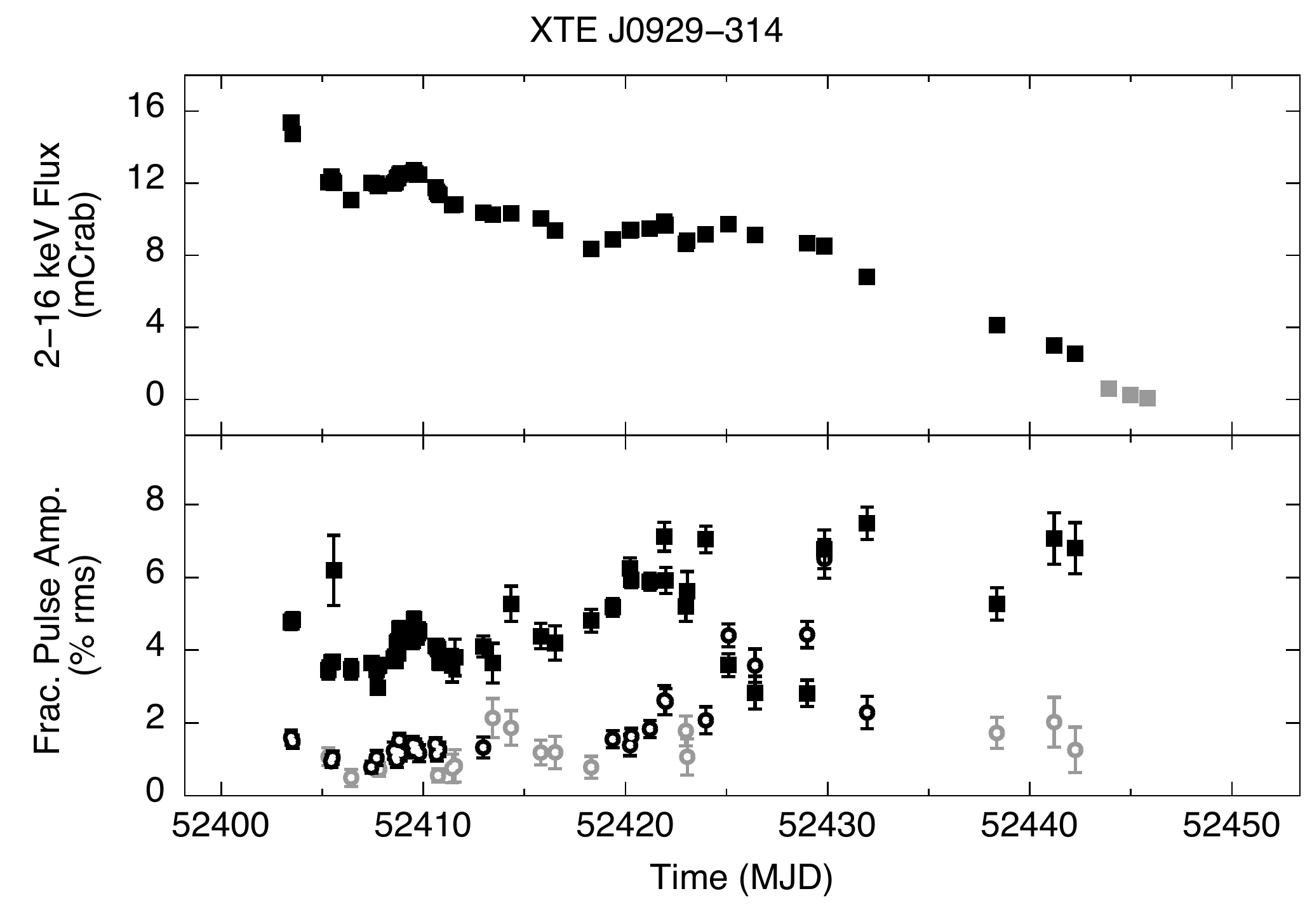}}
	\caption{\small Light curve of the 2002 outburst of XTE J0929--314. Legends are same as in Fig.~\ref{fig.J1756}.}
	\label{fig.J0929}
\end{figure}
XTE~J0929--314 was discovered in April 2002 with {\it RXTE}, and the 185~Hz 
pulsations were immediately detected \citep{remil02b,remil02c}.
The source has been detected in outburst only once, with the light curve shown
in Fig.~\ref{fig.J0929}. %The highest and lowest flux with pulsations are found
% on MJD~52403.5 and 52442.3.

XTE~J0929--314 is significantly away from the galactic plane (galactic coordinates: 
$260.1^\circ, 14.2^\circ$) and shows a low neutral hydrogen column density 
($\sim 7.6 \times 10^{20} \mbox{ cm}^{-2}$, \citealt{juett03}). The background contribution for this
source is therefore negligibly low, and indeed could not be measured as the non-pulsating 
observations at the end of the outburst, which have too few counts to constrain the spectrum. 

There are no good estimates of the distance to the source, with the only constraint 
claiming $d \gtrsim 6$~kpc based on estimates of average accretion rate \citet{gallo02}. 
Using this distance we obtain a magnetic field range of $1.2\E{7}~\mbox{G} < B < 1.2\E{9}$~G.

\subsection{XTE J1807.4--294} \label{sec.res.bkg}
\begin{figure}
	\centering
	\includegraphics[width=\linewidth] {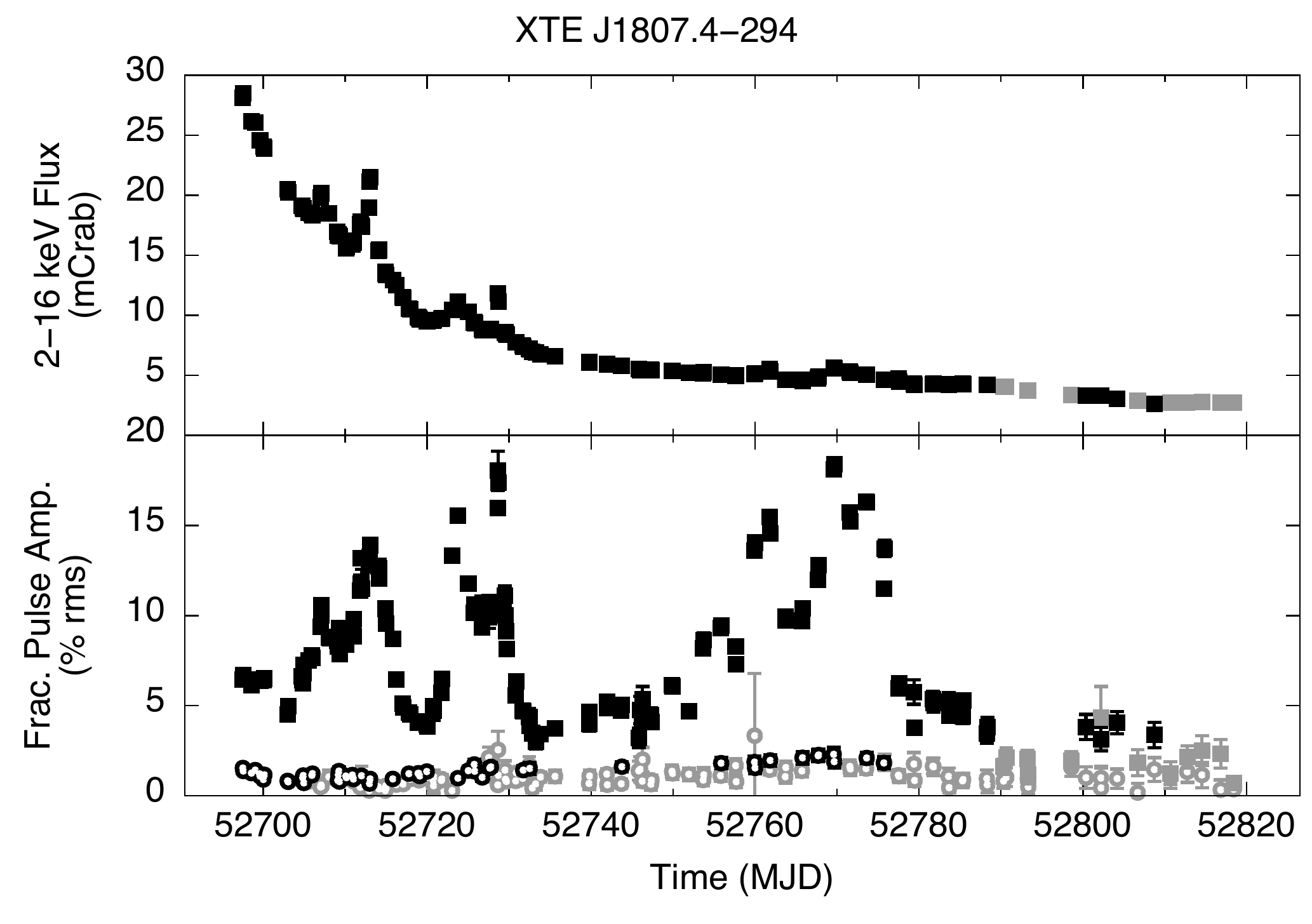}
	\caption{\small Light curve of the 2003 outburst of XTE J1807.4--294. Legends are same as in Fig.~\ref{fig.J1756}.}
	\label{fig.J1807}
\end{figure}
XTE~J1807.4--294 was discovered in February 2003 \citep{markw03b} and the 191~Hz pulsations
were immediately found with the {\it RXTE} observations. 
The source has been in outburst only once. We find the highest and lowest
flux with pulsations to occur on MJD~52697.6 and MJD~52808.7, respectively 
(Fig.~\ref{fig.J1807}). 

We measured the background contribution on MJD~52816.8, but note that this 
background level is similar to the lowest flux with pulsations, such that
the uncertainty in the background estimation is larger than the apparent source
contribution. To be conservative we calculate the upper limit to the 
magnetic moment without adjusting for the background. This implies  that we take the 
low flux observation as an upper limit to the true lowest flux at which pulsations 
are present. If the presence of pulsations can be established at a lower flux level,
the upper limit will decrease, thus tightening the allowed magnetic field range.

There are no well defined estimates for the distance to this source. 
Some authors assume the source is near the galactic centre \citep{falan05,campa03},
and take the distance to be $\sim 8$~kpc. Others, however, estimate the distance 
at $\sim 4.4$~kpc \citep{riggi08}, by comparing the observed flux to the accretion rate
inferred from the pulse timing analysis. 
Lacking a more robust estimation of the distance, we adopt a distance of $4.4$~kpc. 
We then arrive at a magnetic field range estimate of $1.1\E{7}~\mbox{G} < B < 1.9\E{9}$~G.

\subsection{NGC 6440 X-2}
\begin{figure}
	\centering
	\includegraphics[width=\linewidth] {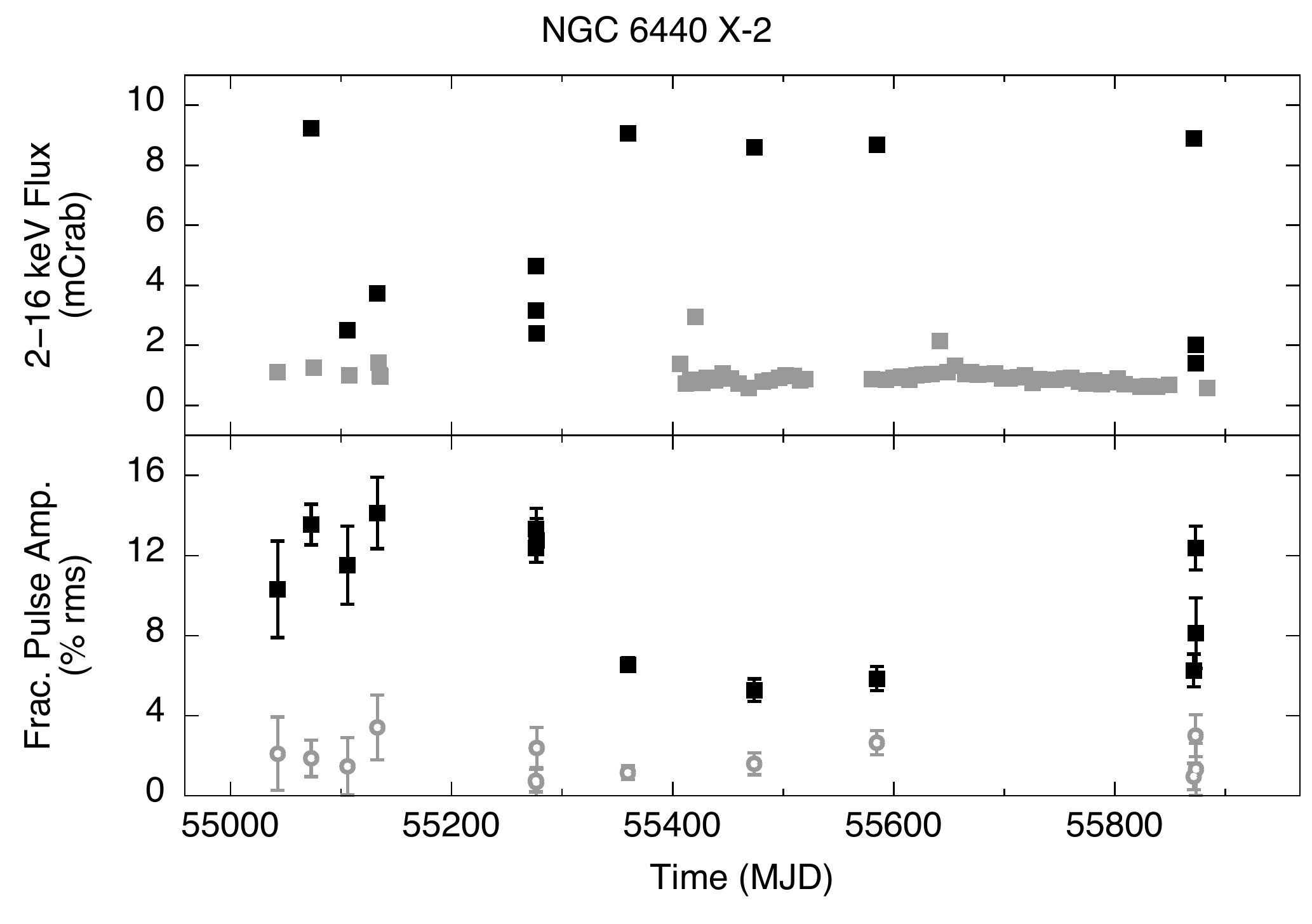}
	\caption{\small Light curve of NGC 6440 X-2 from July, 2009 to November, 2011. Legends are same as in Fig.~\ref{fig.J1756}.}
	\label{fig.ngc6440}
\end{figure}
NGC~6440~X-2 is located in the globular cluster NGC~6440 and was detected serendipitously 
with {\it Chandra} in July 2009 \citep{heink09,heink10}. 
Pulsations at 206~Hz were discovered from subsequent {\it RXTE} observations
\citep{altam09,altam10c}. 

The outburst behaviour of NGC~6440~X-2 is atypical, as it shows brief outbursts of
a few days with a recurrence time as short as one month (see Fig.~\ref{fig.ngc6440}). 
Due to the high-cadence monitoring of this globular cluster, the background level 
is well established between MJD~55700--55850. We estimate the background flux on 
MJD~55823.2.
% The highest and lowest flux levels are found on MJD~55073.1 and~55873.3.

The distance to the cluster NGC 6440 is well constrained to $d = 8.2$~kpc 
\citep{valen07}, and gives a magnetic field range of $1.2\E{7}~\mbox{G} < B < 7.6\E{8}$~G.

\subsection{IGR~J17511--3057}
\begin{figure}
	\centering
	\includegraphics[width=\linewidth] {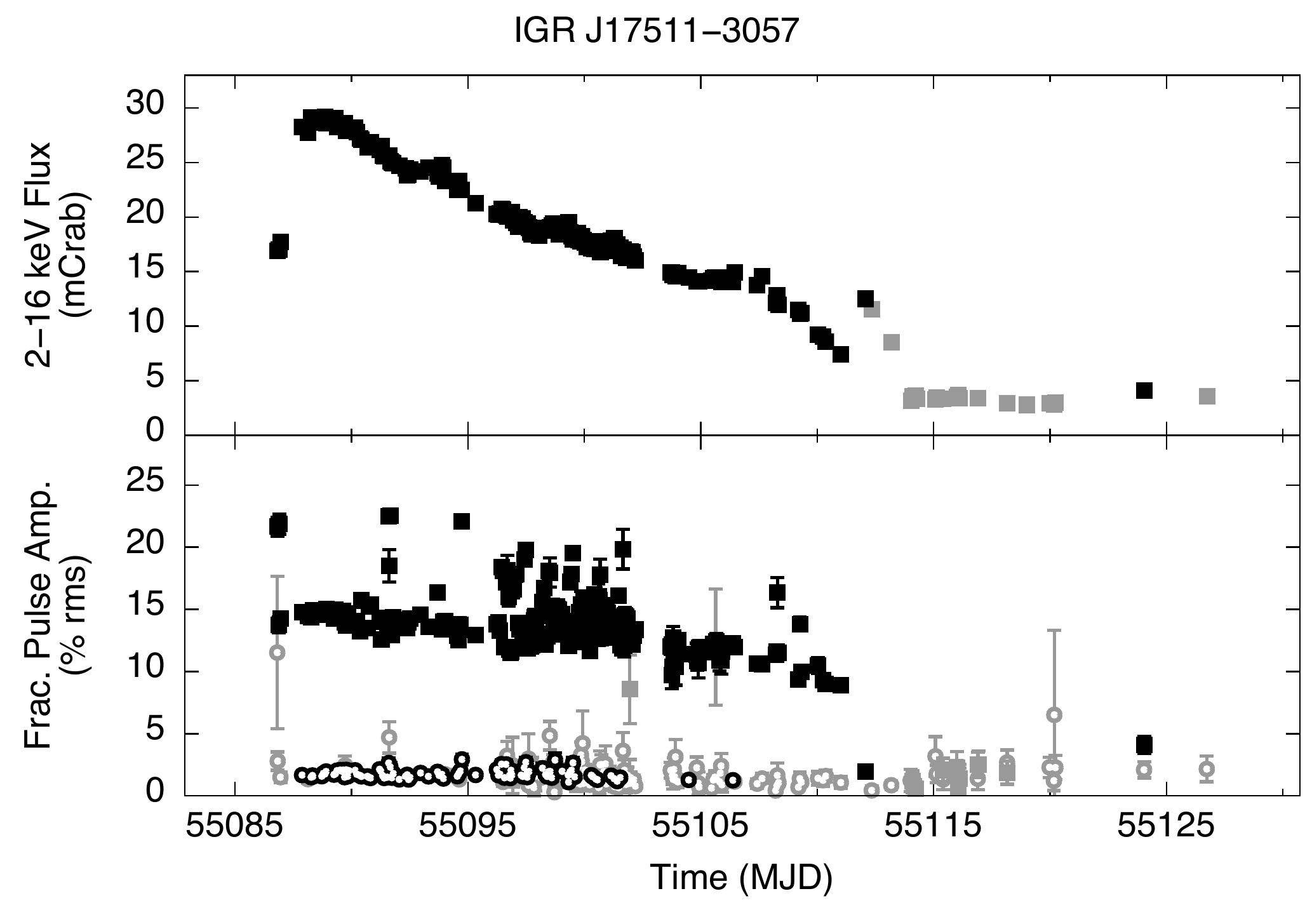}
	\caption{\small Light curve of the 2009 outburst of IGR~J17511--3057. Legends are same as in Fig.~\ref{fig.J1756}.}
	\label{fig.J17511}
\end{figure}
IGR~J17511--3057 was discovered with {\it INTEGRAL} in September 2009 \citep{baldo09},
with subsequent {\it RXTE} observations discovering 245~Hz pulsations 
\citep{markw09}.

The outburst light curve shows a notable flare after MJD~55110 (Fig.~\ref{fig.J17511}) 
which is attributed to simultaneous activity of XTE~J1751--305 \citep{falan11}. 
Pulsations are observed throughout the outburst, and reoccur on MJD~55124.0, 
some 10~days after the source appears to have reached the background level. We select this 
observation as the lowest flux observation with pulsations, and select the observation on MJD~55118.2 
for the background flux. Since the flux difference between these observations is very
small (Table~\ref{table.fluxstates}, we consider the low flux observation to be background
dominated and, like in Sec.~\ref{sec.res.bkg}, neglect the background measurement in calculating
the upper limit on the magnetic moment to obtain a more conservative estimate of the magnetic
field strength.

The distance to this source is estimated at $\lesssim7$~kpc, derived by assuming
the type I X-ray bursts are Eddington limited \citep{altam10b, papit10}. 
Adopting this distance we obtain a magnetic field range of $1.9\E{7}~\mbox{G} < B < 1.2\E{9}$~G.

\subsection{XTE J1814--338}
\begin{figure}
	\centering
	\includegraphics[width=\linewidth] {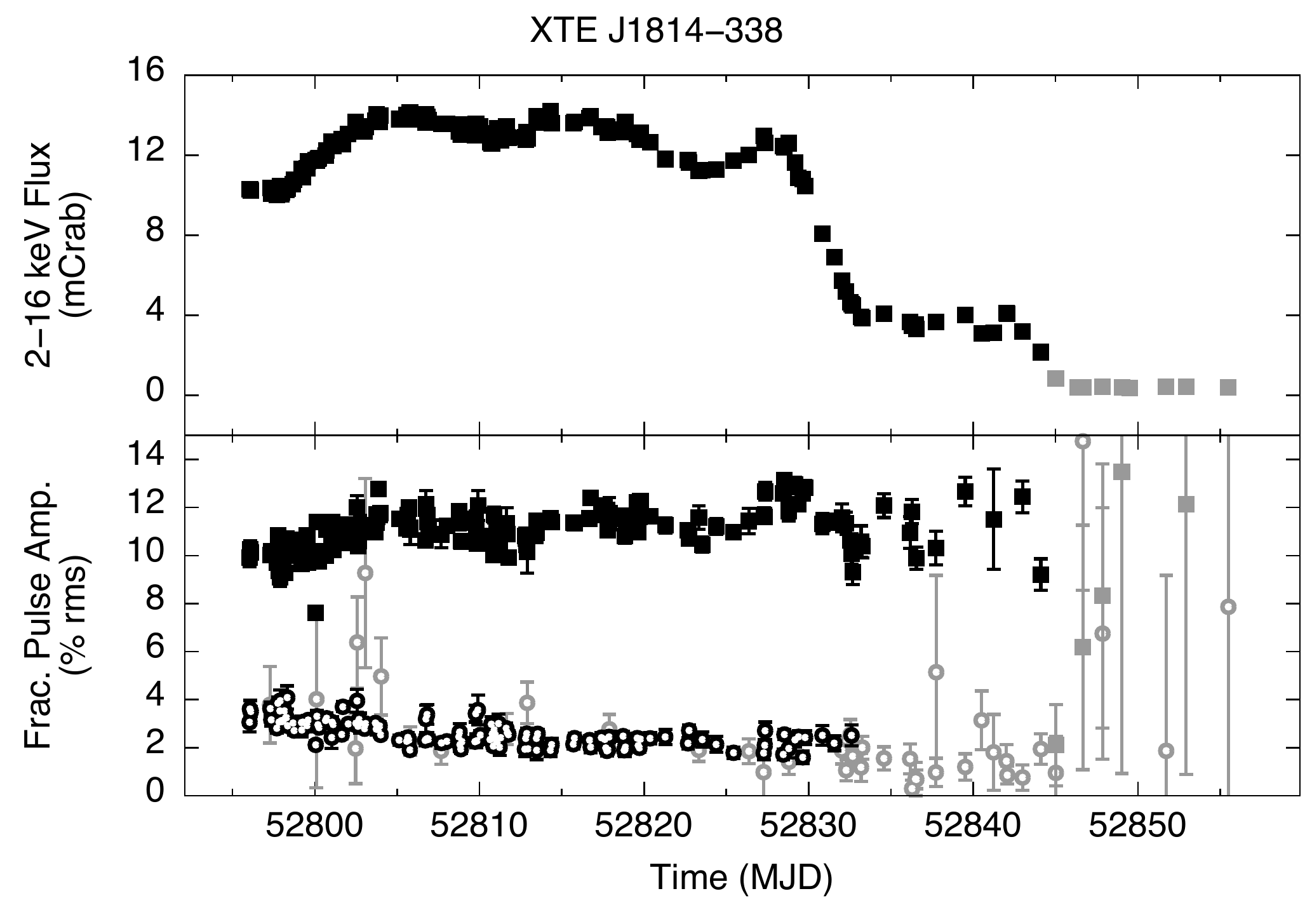}
	\caption{\small Light curve of the 2003 outburst of XTE~J1814--338. Legends are same as in Fig.~\ref{fig.J1756}.}
	\label{fig.J1814}
\end{figure}

XTE~J1814--338 was discovered in June 2003 with {\it RXTE} and immediately
recognized as a 314~Hz pulsar \citep{markw03}.
It has been detected in outburst only once, and shows pulsations
throughout its outburst (Fig.~\ref{fig.J1814}).

% We find the high and low flux levels at MJD~52814.3 and~52844.1.
We measured the background flux from the observation on MJD~52852.9, 
one of the last {\it RXTE} observations of the outburst. 

The distance to the source is estimated at $\lesssim 8$~kpc,  
by assuming the measured luminosity during the type I X-ray burst 
is Eddington limited \citep{stroh03}. 
The resulting magnetic field range is $1.6\E{7}~\mbox{G} < B < 7.8\E{8}$~G.

\subsection{HETE J1900.1--2455}
\begin{figure}
	\centering
	\includegraphics[width=\linewidth] {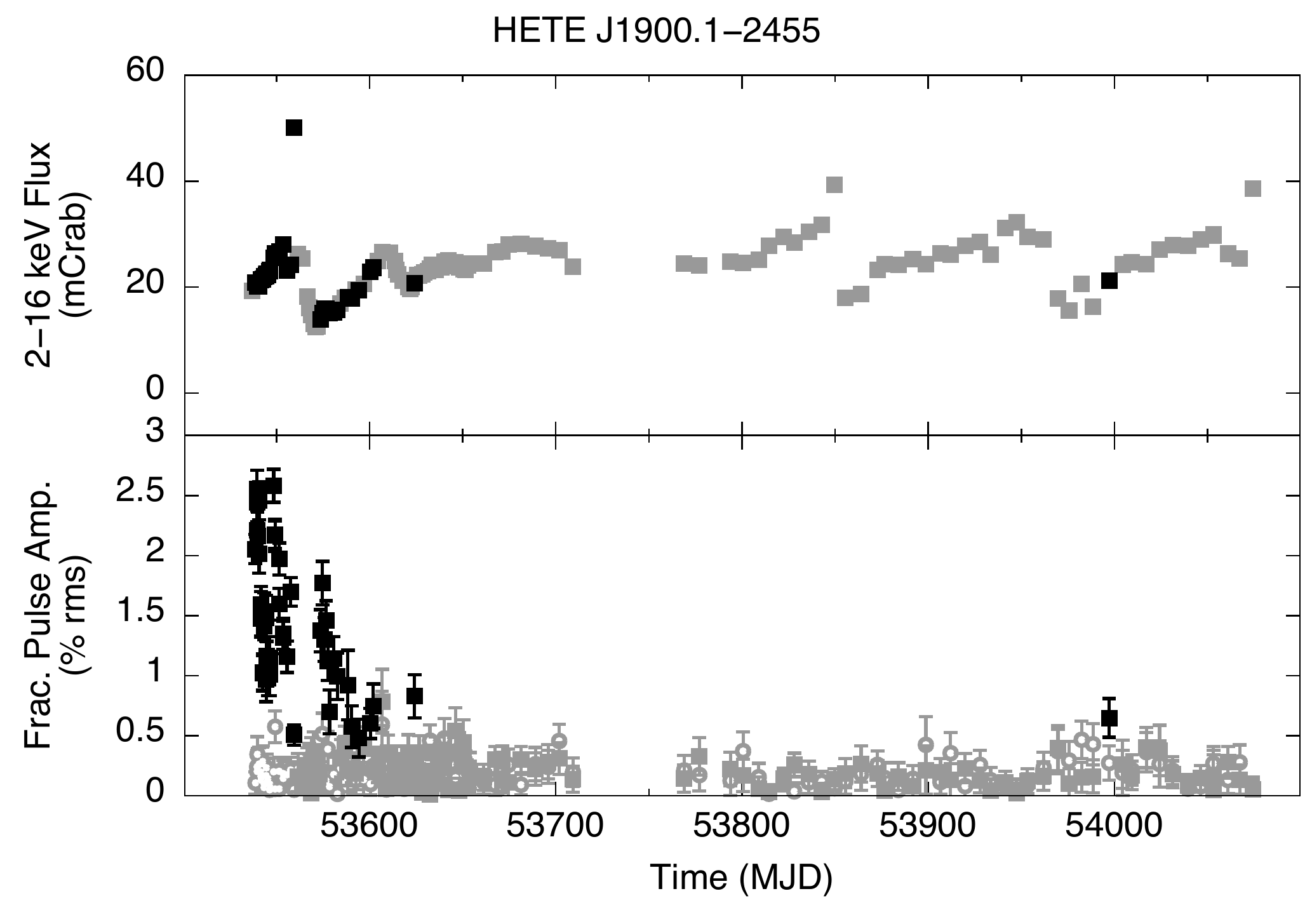}
	\caption{\small 
		Light curve of HETE~J1900.1--2455 from June 2005 to January 2007. Legends are 
		same as in Fig.~\ref{fig.J1756}. Note that there are additional pulse detections 
		beyond $\sim54000$, however, these are tentative \citep{patru12} and do not 
		change the measured flux range. }
	\label{fig.hete}
\end{figure}
HETE~J1900.1--2455 was discovered through a bright type I X-ray burst observed
with the {\it High Energy Transient Explorer 2 (HETE-2)} in June 2005 \citep{vande05}
and 377~Hz pulsations were observed with {\it RXTE} quickly thereafter \citep{morga05,kaare06}.
Unlike the other AMXPs, which show outbursts that last for weeks to months, 
HETE~J1900.1--2455 has been active since discovery and is yet to return to quiescence.
Persistent pulsations have been reported to occur during the first 20 days of the
outburst, after which only intermittent pulsations have been seen \citep{patru12}.

We have analysed all archival {\it RXTE} observations of this source
and find that the observations of highest and lowest flux with pulsations
are on MJD~53559.5 and MJD~53573.8, respectively. 

Since the source has shown continuous activity since its discovery, there
is no observation during quiescence from which the background flux can be
estimated. However, as the source is well away from the galactic centre 
(galactic coordinates: $l=11.3^\circ ,b=-12.9^\circ $), the the background flux
is expected to be comparatively small. We therefore neglect the background contribution 
for this source, noting again that this leads us to slightly overestimate the source flux,
which yields a less stringent constraint on $\mu_{\rm max}$. 

The distance to the source is taken to be $d \sim 5$~kpc, based on photospheric 
radius expansion of a type I X-ray burst \citep{kawai05}. The magnetic field
range obtained is $1.6\E{7}~\mbox{G} < B < 1.0\E{9}$~G.

\subsection{SAX J1808.4--3658}
\begin{figure}
	\centering
	\includegraphics[width=\linewidth] {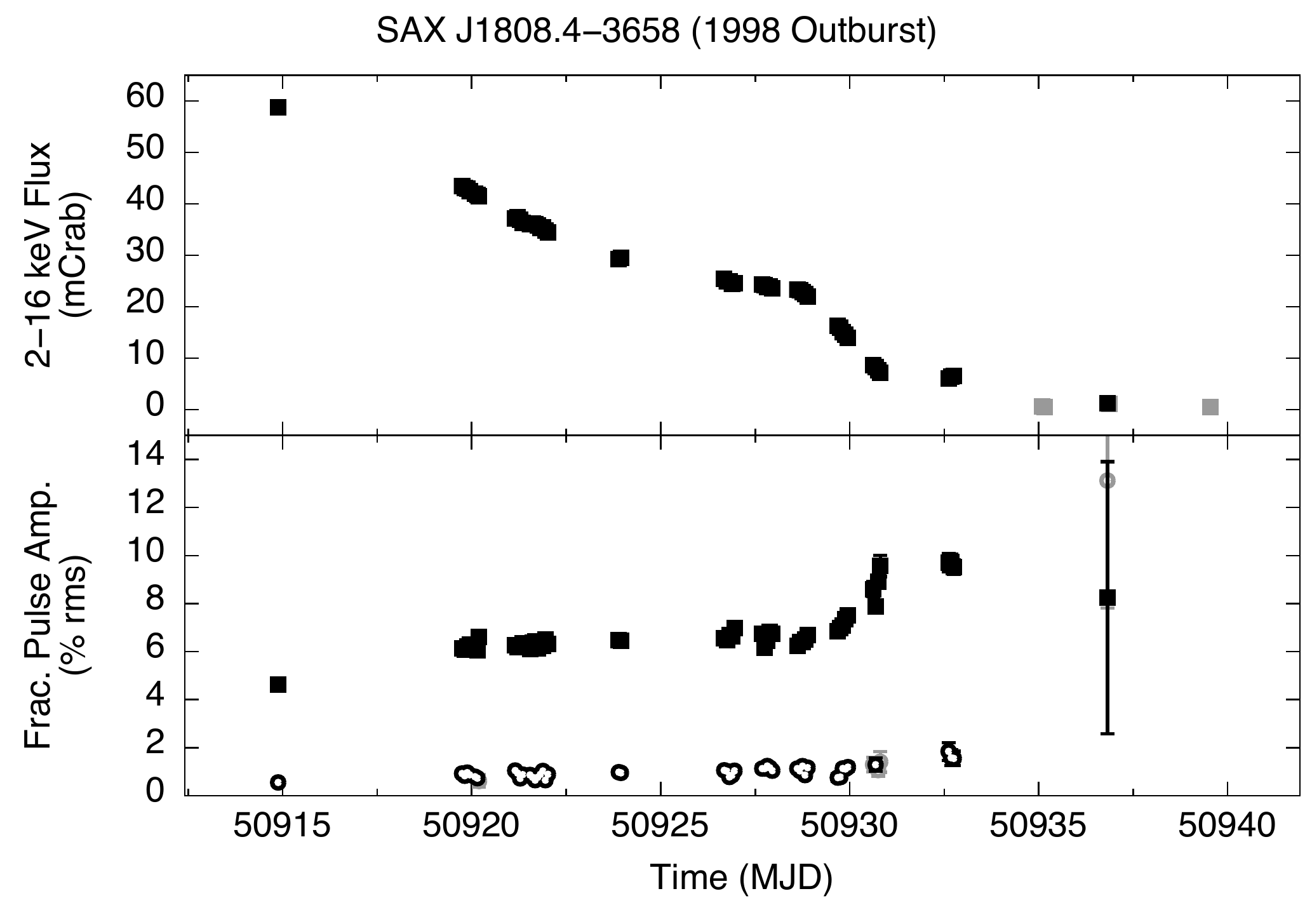}
	\includegraphics[width=\linewidth] {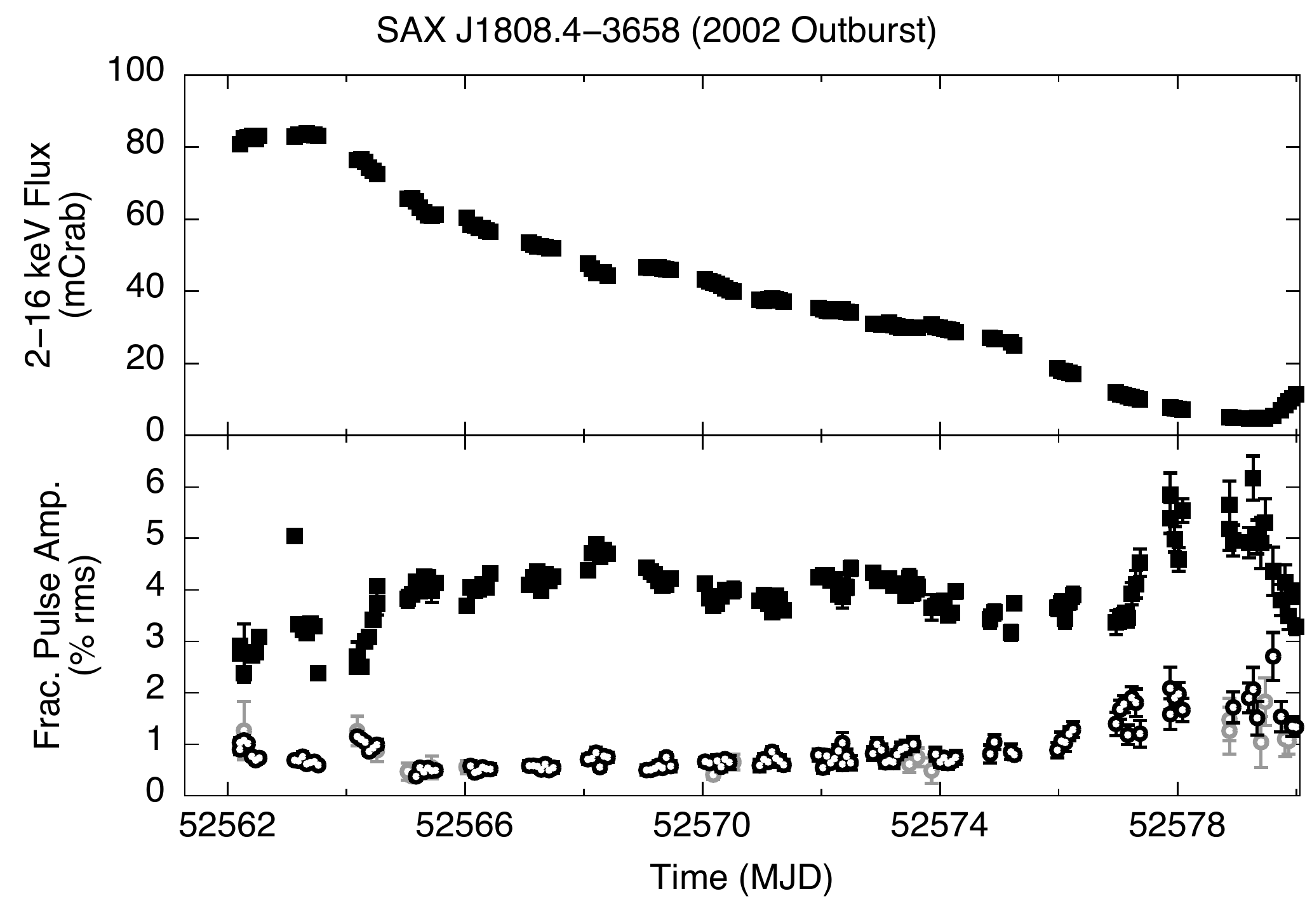}
	\caption{\small Light curve of the 1998 and 2002 outbursts of SAX~J1808.4--3658. Legends are same as in Fig.~\ref{fig.J1756}.}
	\label{fig.J1808}
\end{figure}
SAX~J1808.4--3658 was discovered with {\it BeppoSax} in in 1996 \citep{intza98} 
and the detection of 401~Hz pulsations in 1998 with {\it RXTE} made it the first
known AMXP \citep{wijna98}. SAX~J1808.4--3658 has been observed in outbursts with 
{\it RXTE} six times.
 
We find the highest flux with pulsations to be in the 2002 outburst on MJD~52563.2 and 
the lowest flux with pulsations in the 1998 outburst on MJD~50936.8. The lowest
flux with pulsations is observed toward the end of the 1998 outburst 
(Fig.~\ref{fig.J1808}). Although the flux of this observation is only slightly higher
than that of the background observations, the pulse detection is very significant
and yields a phase that is consistent with the timing model of this outburst.
If pulsations are still present at the same amplitude in the other low flux observations, 
we would not be able to detect them due to the low count rate. We
therefore consider the low flux observation with pulsations to be background dominated.

The distance to SAX~J1808.4--3658 is $d = 3.5$~kpc, and was derived
from photospheric radius expansion in type I X-ray bursts \citet{gallo06}. With
this distance we obtain a magnetic field range of $1.4\E{7}~\mbox{G} < B < 1.8\E{8}$~G.

\subsection{IGR~J17498--2921}
\begin{figure}
	\centering
	\includegraphics[width=\linewidth] {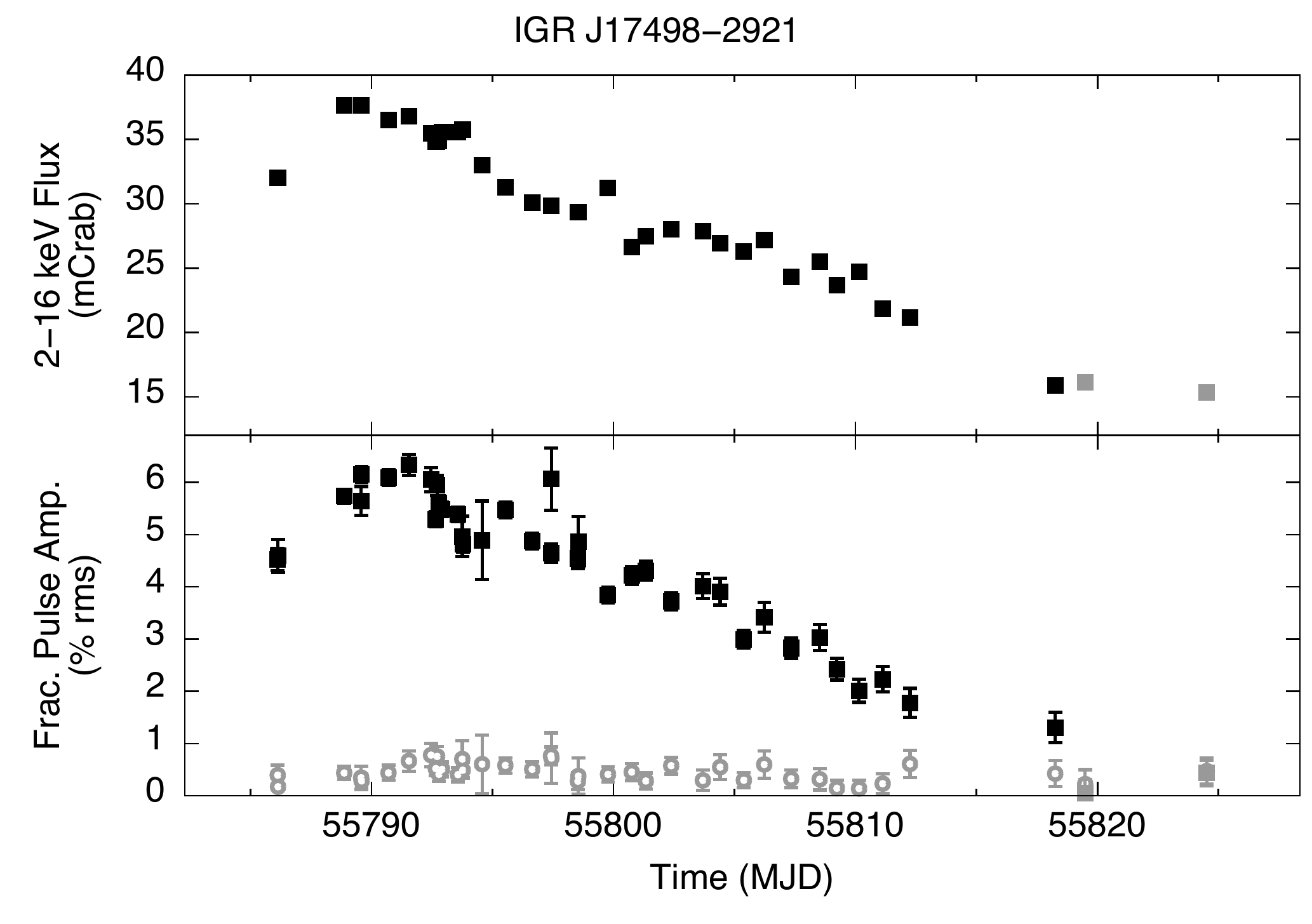}
	\caption{\small Light curve of the 2011 outburst of IGR~J17498--2921. Legends are same as in Fig.~\ref{fig.J1756}.}
	\label{fig.J17498}
\end{figure}

IGR~J17498--2921 was discovered in August 2011 with {\it INTEGRAL} \citep{gibau11}, 
following which pulsations at $401$ Hz were reported by \citet{papit11}. 
The light curve of this outburst is shown in Fig.~\ref{fig.J17498}. 

Since the source is close to the galactic centre (galactic coordinates: $l=0.16^\circ,
b=-1^\circ$), the observations have a large X-ray background contamination, with
the lowest observed flux with pulsations again background dominated.
%The highest and lowest flux with pulsations are found at MJD~55789.6 and~55818.3.

The distance to IGR~J17498--2921 is estimated at $\sim 8$~kpc, based
on photospheric radius expansion during a type I X-ray burst \citep{falan12}. 
We then find a magnetic field range of $2.0\E{7}~\mbox{G} < B < 1.6\E{9}$~G.

%As shown in Fig.~\ref{fig.J17498} the states with and without pulsations are
%not clearly distinguishable. The flux of the last {\it RXTE} observation is
%only $\sim 4\%$ lower than the lowest flux state with pulsations. It is
%possible that although significant pulsations were not detected, the obtained
%flux can be from residual weak accretion. Overestimation of the galactic ridge
%emission may lead to erroneous tighter constraints on the magnetic field
%limits, as discussed later in Sec.~\ref{sec.discuss}.

\subsection{XTE J1751--305}
\begin{figure}
	\centering
	\includegraphics[width=\linewidth] {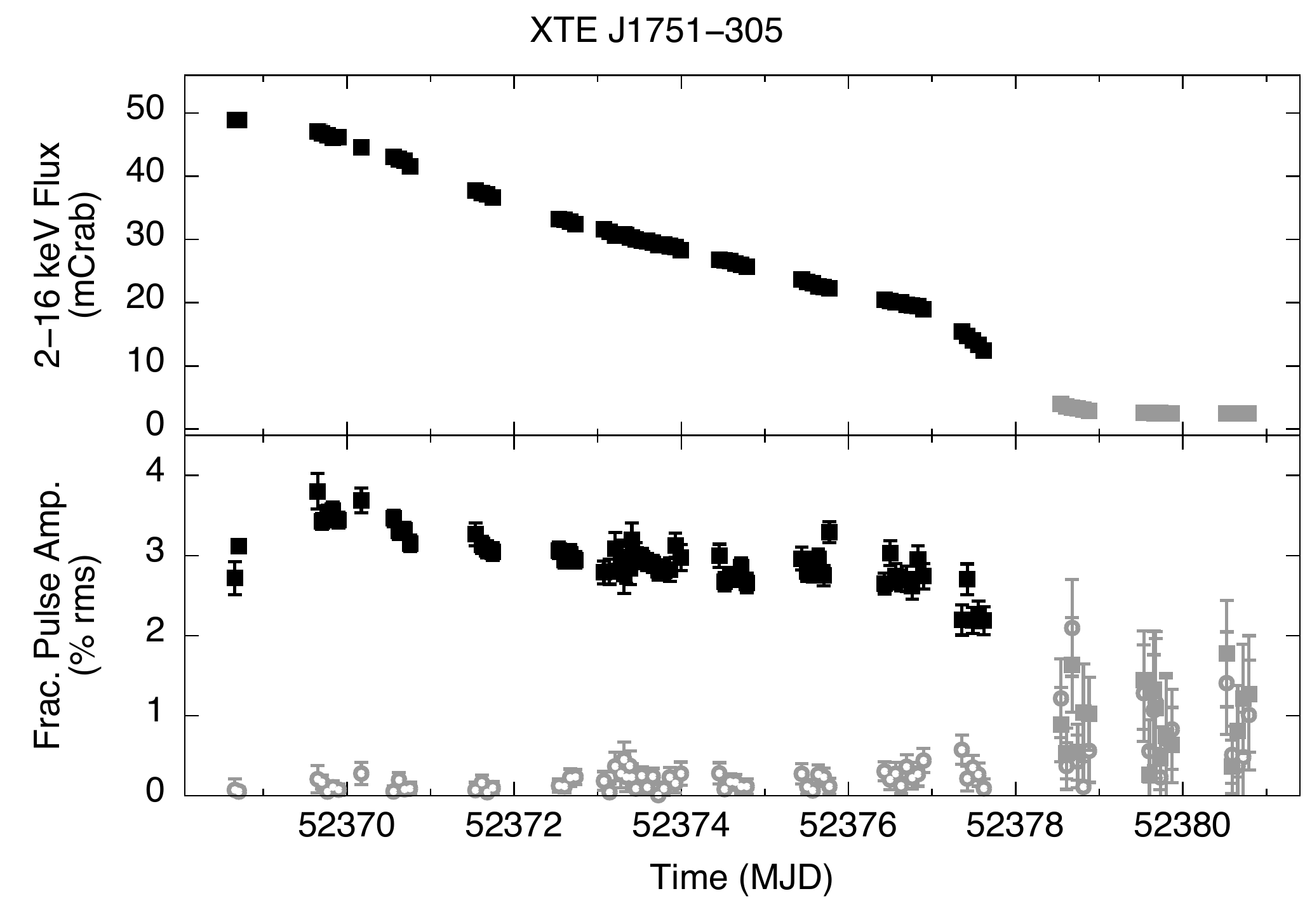}
	\caption{\small  Light curve of the 2002 outburst of XTE~J751--305. Legends are same as in Fig.~\ref{fig.J1756}.}
	\label{fig.J1751}
\end{figure}
XTE~J1751--305 and its 435~Hz pulsations were discovered with {\it RXTE} in April 2002
\citep{markw02}.  A brief second outburst was detected in 2009 \citep{markw09},
which was coincident with on-going activity of IGR~J17511--3057 in the same
field of view. Since {\it RXTE} is not an imaging detector, the flux
contribution of these two sources cannot be separated, so we restricted our
analysis to the 2002 outburst of XTE~J1751--305 only.

%We find the the highest and lowest flux level with pulsations to occur at MJD~52368.7 and~52377.5
We estimate the background from the last {\it RXTE} observation on MJD~52380.7, 
when the pulsations were no longer detected (see Fig.~\ref{fig.J1751}).

The distance to this source is not well-defined. \citet{markw02} constrain 
the distance to $\gtrsim 7$~kpc by equating predicted mass transfer rates 
\citep{rappa83,king97} to values inferred from X-ray observations.
\citet{papit08} instead compare the spin-frequency derivative to models for
angular momentum exchange \citep{rappa04}, and so constrain the distance to 
$6.7-8.5$~kpc. For our analysis, we take a distance of 7~kpc, which
results in a magnetic field range of $2.5\E{7}~\mbox{G} < B < 1.1\E{9}$~G.

\subsection{SAX J1748.9--2021}
\begin{figure}
	\centering
	\includegraphics[width=\linewidth]{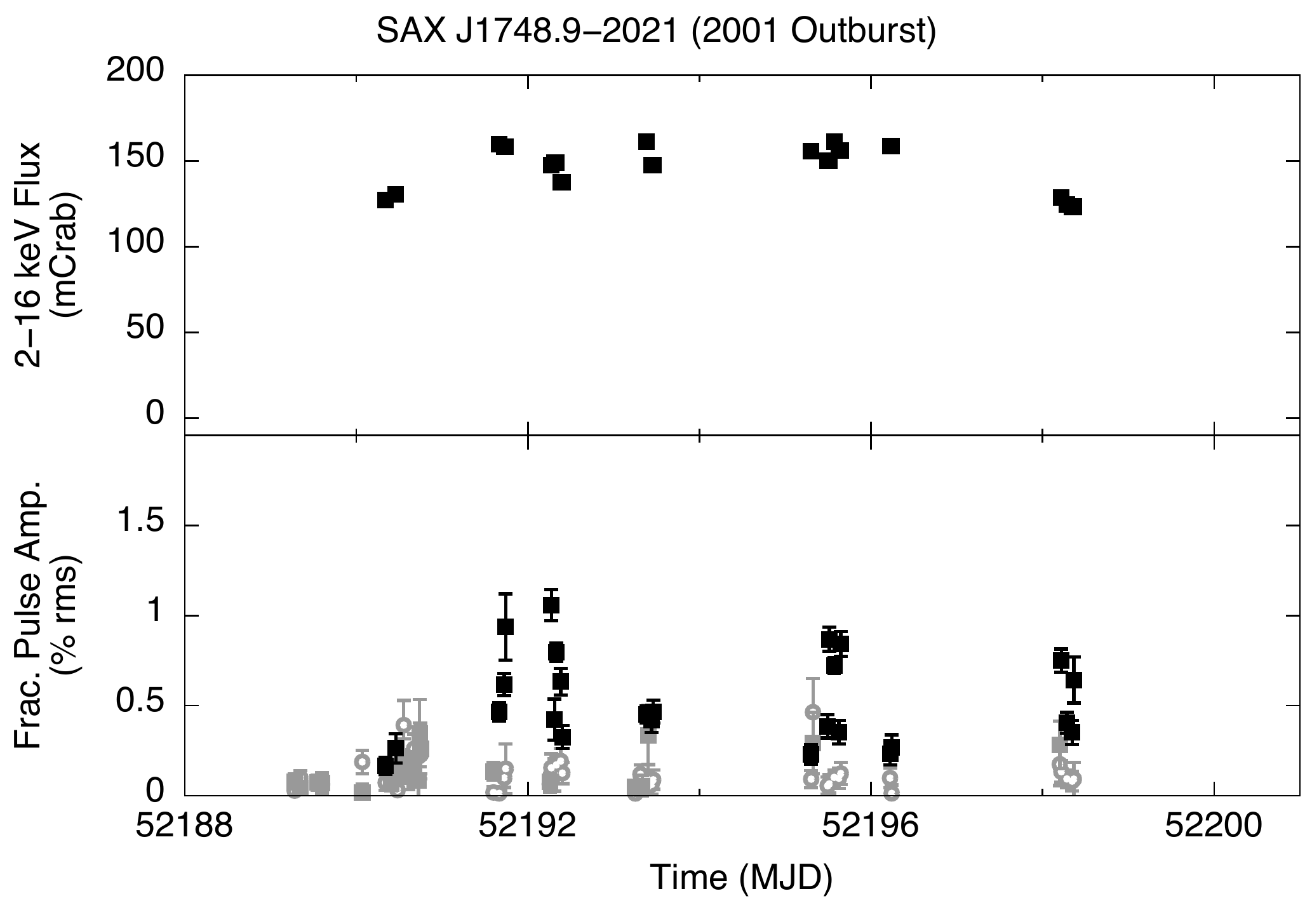}
	\includegraphics[width=\linewidth]{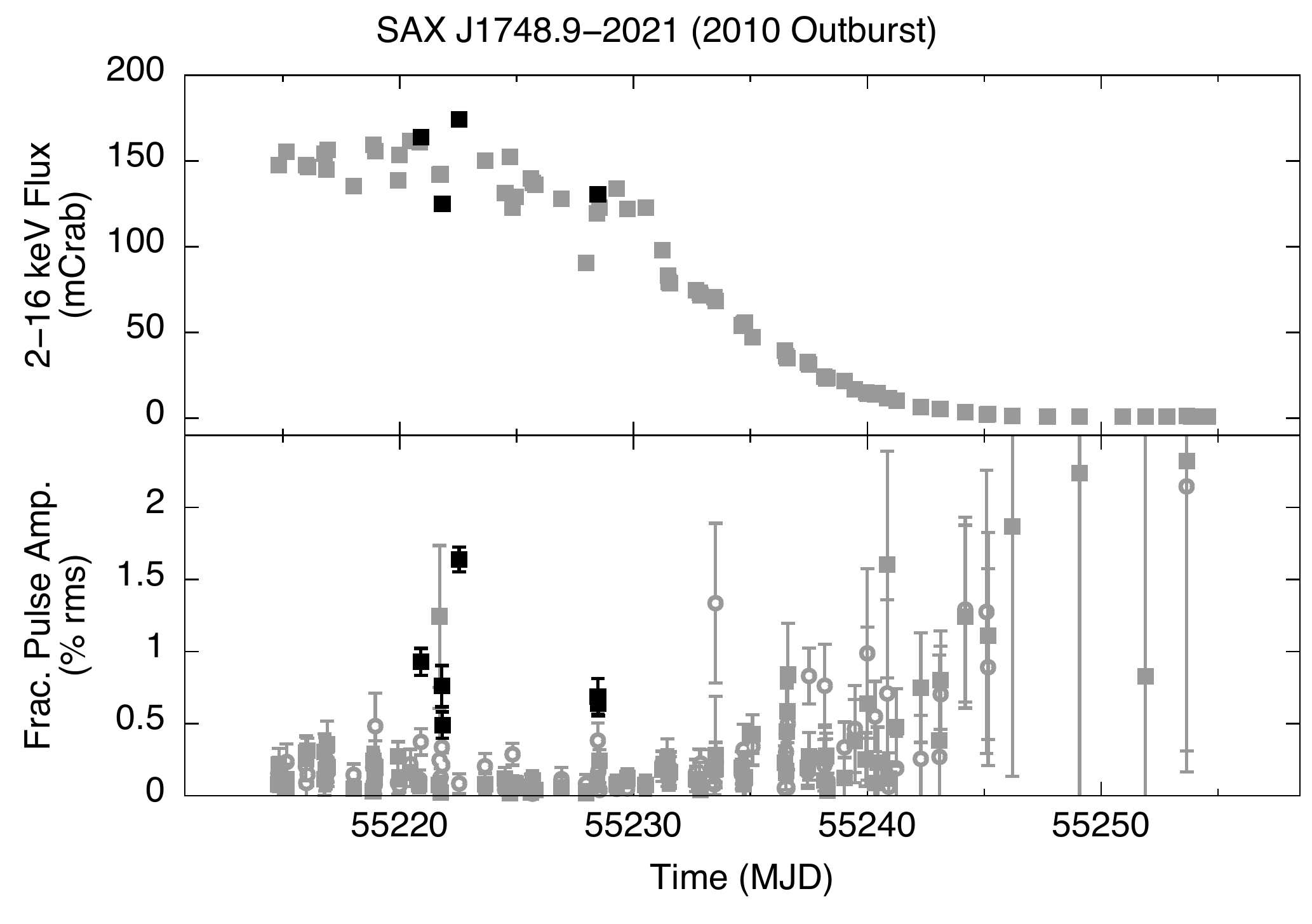}
	\caption{\small Light curves of the 2001 and 2010 outbursts of SAX J1748.9--2021. 
		Legends are same as in Fig.~\ref{fig.J1756}. }
	\label{fig.J1748}
\end{figure}

SAX~J1748.9--2021 is located in the globular cluster NGC~6440
\citep{intza99}, and was observed in outburst by {\it RXTE}
in 1998, 2001, 2005 and 2010 \citep{intza99, intza01, mark05, patru10a}.
Among these outbursts 442~Hz intermittent pulsations have been detected in 2001, 2005 and
2010 \citep{gavri07,altam08,patru10a}.

We find that the highest flux with pulsations occurs during the 2010 outburst on MJD~55222.5, 
while the lowest flux with pulsations is seen in the 2001 outburst on MJD~52198.3 
(light curves shown in Fig.~\ref{fig.J1748}).

As the source is associated with a globular cluster, its distance is comparatively
well constrained to $8.2$~kpc \citep{valen07}. We find a magnetic
field range of $4.9\E{7}~\mbox{G} < B < 3.8\E{9}$~G.

\subsection{Swift J1749.4--2807}
\begin{figure}
	\centering
	\includegraphics[width=\linewidth] {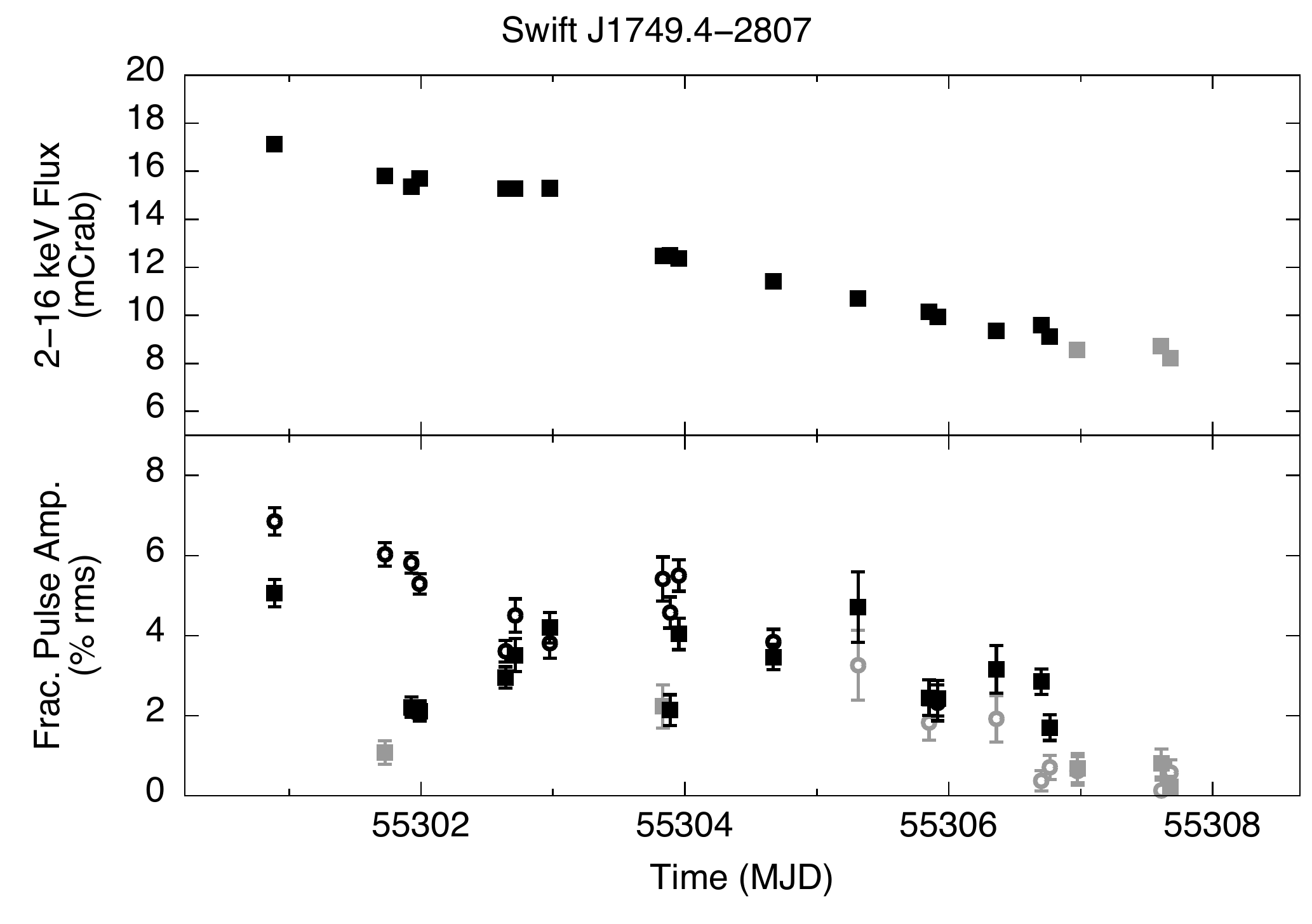}
	\caption{\small Light curve of the 2010 outburst of Swift J1749.4--2807. Legends are same as in Fig.~\ref{fig.J1756}. }
	\label{fig.J1749}
\end{figure}
Swift J1749.4--2807 was first detected in June 2006 \citep{schad06},
but its 518~Hz pulsations were not found until the second outburst in April 2010
\citep{altam10a,altam11a}. We find both the highest and lowest flux with 
pulsations occur in this second outburst on MJD~55300.9 and~55306.7, respectively.

The source is located close to the galactic centre and has a strong contaminating
X-ray background flux. The absence of pulsations (Fig.~\ref{fig.J1749}) indicate that
the source was no longer accreting for the last three {\it RXTE} observations 
($>$~MJD~55307), which is confirmed by a source non-detection with both {\it Swift} and {\it INTEGRAL}
\citep{ferri11}. We therefore use the last {\it RXTE} observation to measure the background
flux. As this background flux is comparable to the lowest observed flux with pulsations,
we cannot confidently estimate the source contribution and to be conservative we again calculate 
the upper limit on the magnetic field without using the background.

The distance to the source is $d = 6.7\pm1.3$~kpc, which was inferred from the 
luminosity of a suspected type I X-ray burst \citep{wijna09}. Adopting the central
value for the distance and the measured flux we obtain a magnetic field range of 
$1.1\E{7}~\mbox{G} < B < 7.7\E{8}$~G.

\subsection{Aql X-1}
In 18 outbursts across $\sim 15$~years, Aql~X-1 has shown its 550~Hz pulsations
only for a single episode of about 150~seconds \citep{casel08,messe14}. Hence we cannot
measure a flux range for the presence of pulsations for this source. Instead
we calculate both lower and upper limits on the magnetic field from the same
measured flux, thus obtaining a very conservative estimate of the allowed magnetic
field range. 

The distance to the source is $4.4-5.9$~kpc and was obtained from a photospheric 
radius expansion during a type I X-ray \citep{jonke04b}. 
For a distance of 5 kpc and the measured flux we obtain a magnetic field range of 
$4.4\E{7}~\mbox{ G} < B < 3.1\times 10^9$~G.

\subsection{IGR~J00291+5934}
\begin{figure}
	\centering
	\includegraphics[width=\linewidth]{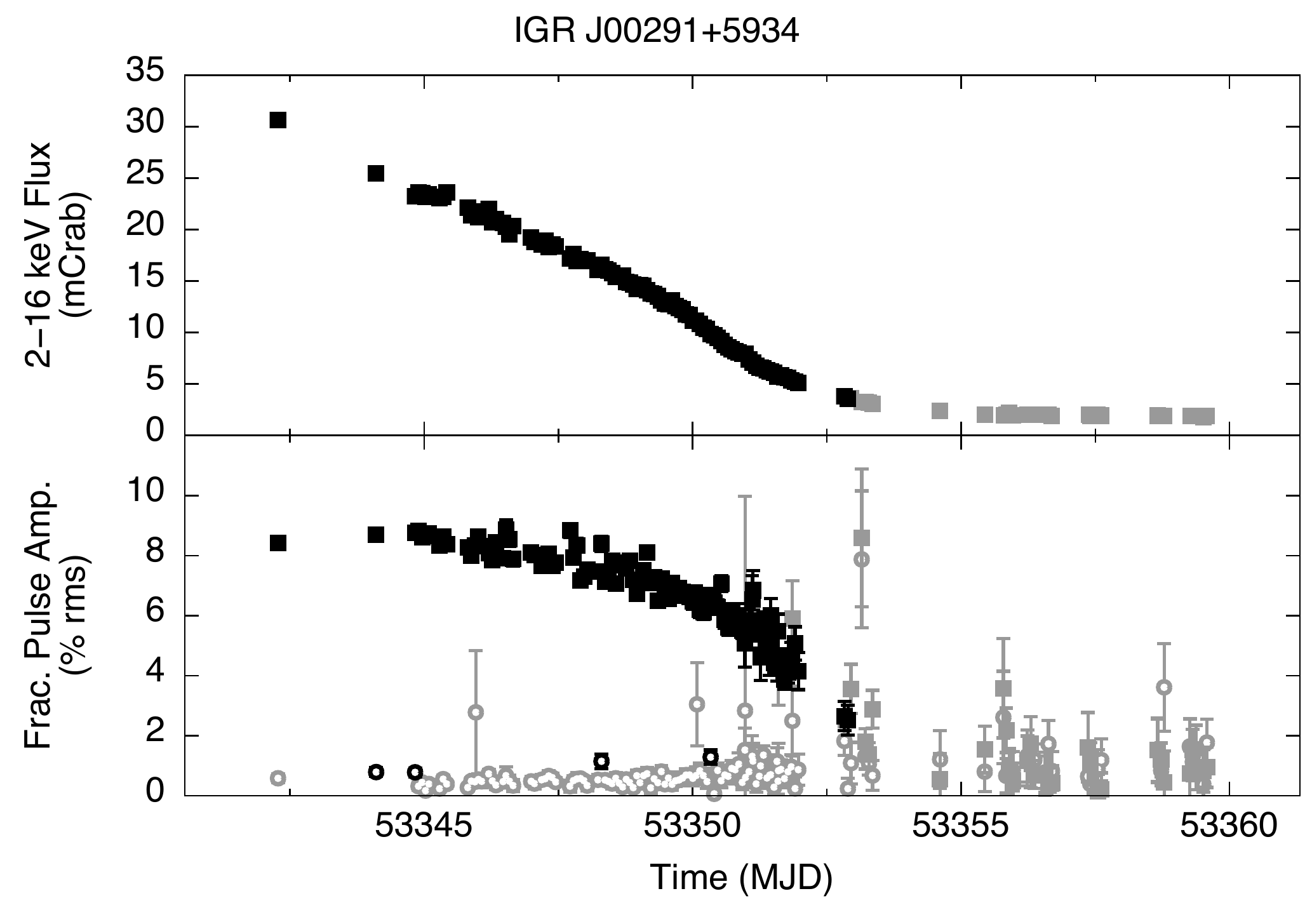}
	\caption{\small 
        Upper panel: Light curve of the 2004 outburst of IGR~J00291+5934. Legends are same as in Fig.~\ref{fig.J1756}. }
	\label{fig.J00291}
\end{figure}
IGR~J00291+5934 was discovered with {\it INTEGRAL} in December 2004 \citep{ecker04,
shaw05} and its 599~Hz pulsations were detected in follow-up {\it RXTE} observations
\citep{markw04}. Outbursts were detected again in August and September 2008
\citep{chakr08,lewis08}. We find both the highest and the lowest flux with pulsations 
occur in the 2004 outburst. % at MJD~53342.3 and~53352.9, respectively. 

It is evident from the light curve (Fig.~\ref{fig.J00291}) that the source 
gradually decays to the background level, which is due to an
intermediate polar in the same field of view \citep{falan05b}.
%However, at the low flux end, it is not clear whether the pulsations cease of if
%the detection is limited by the decreasing signal-to-noise. 
Because the lowest flux with pulsations is again comparable to the estimated background flux,
we assume the low flux observation is background dominated to get a more
conservative magnetic field estimate.

The distance to IGR~J00291+5934 has been estimated in several ways.
From the long term average accretion rate, \citet{gallo05} constrain the 
distance to $\lesssim 4$~kpc.
Comparing the observed quiescent flux to that of SAX~J1808.4--3658,
\citet{jonke05} estimate the distance to be $2-3.6$~kpc. 
Similar estimates were also obtained by \citet{torre08}, who
report $d = 1.8-3.8$~kpc by modeling the light curve.
In this work we adopt the central distance of $d = 3$~kpc, which gives a 
magnetic field range of $8.5 \times 10^6~\mbox{G} < B < 1.9 \times 10^8$~G.

\section{Discussion}\label{sec.discuss}
In this section we discuss the various sources of uncertainty involved in the
magnetic field estimates presented in this work.
\begin{figure}
	\centering
	\includegraphics[width=\linewidth]{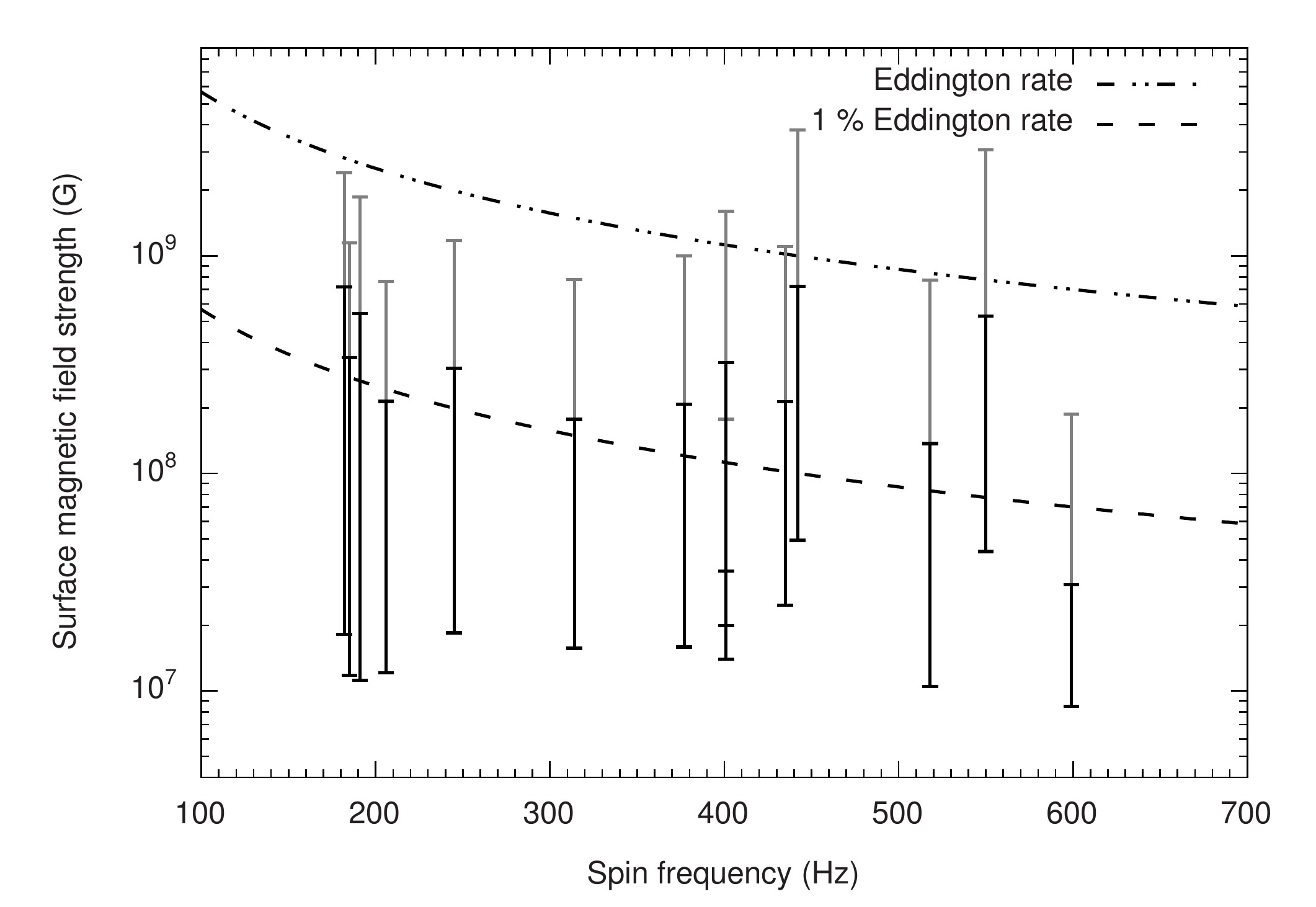}	
	\caption[B-P plot of AMXPs]{\small Magnetic field vs spin frequency for AMXPs analysed in this work. The grey lines show the range of estimated field strengths of the sources from Table.~\ref{table.fieldvalues}. The black lines represent the revised upper limits following eq.~\ref{eq.truncR} which gives tighter constraints of the field strength estimates. The dashed and dashed-dotted curves show the equilibrium spin rate for two different accretion rates.}
	\label{fig.amxpbp}
\end{figure}

%\begin{enumerate}

%\item{\it Masses and radii of AMXPs:}
\subsection{Distance estimates}
    The largest uncertainties in our field estimates are due to poor constraints on the
    distance. Errors in the distance affect the luminosity estimates and cause a 
    systematic uncertainty that scales both the upper and lower limit by the same 
    factor ($B_{\rm min/max} \propto d$). The sources associated with globular clusters
    are not greatly affected by this uncertainty as they have well measured distances. Some 
    AMXPs exhibit thermonuclear bursts with  photospheric radius expansion, which allows for 
    tight constraints on the distance. For the remaining sources, however the distance measurements 
    are less certain and the resulting systematic uncertainty has a more prominent role.
    In Table~\ref{table.fieldvalues} we explicitly report the distances used in calculating
    the upper and lower limits, such that our estimates can be easily adjusted should
    a more accurate measure distance be obtained for one of these sources.

\subsection{Masses and radii of AMXPs}
    Currently there are no reliable estimates available for the masses or radii of
    AMXPs. In calculating the field strengths presented in
    Table~\ref{table.fieldvalues} we assumed the neutron stars to be of canonical
    mass $M \sim 1.4 M_\odot$ and radius $\sim 10$~km. However, theoretical
    calculations of mass-radii relations \citep[e.g.][]{latti01,latti14} predict 
    that both parameters may vary over a wide range of $M\sim 0.2-2.9 M_\odot$ 
    and $R\sim 10-15$ km.
    
    For a more massive neutron star of $\sim2M_\odot$ the upper limit on the
    magnetic dipole moment will increase by $\sim 12.6\%$ ($\mu_{\rm max} \propto
    M^{1/3}$), whereas the lower limit will decrease by $8.2\%$ ($\mu_{\rm min}
    \propto M^{-1/4}$) from the values reported in Table~\ref{table.fieldvalues},
    resulting in a broadening of the estimated range.
    
    For a neutron star of larger radius, e.g. $\sim 15$~km, the upper limit on the
    magnetic dipole moment will scale as $\mu_{\rm max} \propto R^{1/2}$, causing
    an increase of $22.5\%$ with respect to the values we report. The lower limit
    to the magnetic dipole moment scales with radius as $\mu_{\rm min} \propto
    R^{9/4}$, such that assuming a 15 km radius increases the lower limit by
    $149\%$. For a $R=15$~km neutron star we then find that the allowed range of
    magnetic field strength moves to higher values and covers a slightly broader
    range\footnote{The field strength range covers more than an order of magnitude,
	so despite the large fractional change, the absolute shift of the lower
    limit is smaller than that of the upper limit.}.

\subsection{Disk-magnetosphere interaction} \label{sec.bdiskinteraction}
	In this paper we made the assumption that the neutron star magnetic field is dipolar
    in nature, which may not be valid. Near the truncation radius the pressure 
    the disk exerts on the magnetosphere can affect the field geometry, and near
    the neutron star surface this may be further complicated by higher multipole moments 
    that could dominate over the dipole component.
    We parametrized the effect of the field shape with the factor $\gamma _B$ (see eq.~\ref{eq.trunc}),
    which we assumed to vary over a wide range of $0.01-1$ to account for such uncertainties. 
    This range is consistent with the results of numerical simulations, which suggest the largest
    range of truncation radii is $r_t \sim(0.5-1.2) r_A$ \citep{zanni09,roman08}, implying 
    $\gamma_B \simeq 0.06-1.3$.

\begin{itemize}
	\item
    Recent MHD simulations \citep{zanni13,kulka13,long08,roman08} have confirmed that 
    complex field topologies arise at the inner edge of the disk. \citet{kulka13} show 
    that if the truncation radius is in the range $2.5 R_s < r_t < 5 R_s$, the
    non-dipolar field structure results in a modified expression of the
    truncation radius 
    \begin{equation}\label{eq.truncR}
    r_t = 1.06 R_s \left(\frac{\mu ^4}{G M \dot{M}^2 R_s^7}\right)^{1/10}
    \end{equation}
    This modification is relevant when estimating the upper limit to the magnetic
    field, for which we take the disk to be truncated at the co-rotation radius,
    which normally falls within the range of radii where eq.~\ref{eq.truncR} is
    applicable. Using this expression for the truncation radius we obtain a
    modified boundary layer parameter as
    \begin{equation}
    \widetilde{\gamma}_B = 0.0616 \left(\frac{\nu _s}{100 {\rm~Hz}}\right) 
    	\left(\frac{M}{1.4~M_\odot}\right)^{-1/2} 
		\left(\frac{R_s}{10{\rm~km}}\right)^{3/2} \label{eq.newgamma}
    \end{equation}
    which, instead of conservatively assuming 0.01, may be used in eq.~\ref{eq.mumax} 
    to obtain a more constraining upper limit on the magnetic dipole moment. 

	\item
    Near the neutron star surface higher multipole moments of the magnetic field 
    may be stronger than the dipole component. As shown in numerical studies 
    \citep{long07, long08} such complex non-dipolar field configurations strongly 
    affect the inner accretion geometry, but the effect on $\gamma_B$ is not
    well established. If further theoretical considerations can constrain
    $\gamma_B$ to a smaller value than our conservative assumption of $\gamma _B = 1$, 
    then this could be used to tighten the lower limit on the magnetic field 
    strength ($B_{\rm min} \propto \gamma _B^{-1/2}$).
\end{itemize}
    
    In Fig.~\ref{fig.amxpbp} we plot the conservative ranges of the surface field
    strength following the analysis in Sec.~\ref{sec.diskinteraction} in grey and
    present the more constraining estimates based on eq.~\ref{eq.newgamma} in
    black. The dashed lines represent the equilibrium spin (Alfv\'en radius at
    co-rotation) for a mean long term accretion rate
    \begin{align}
        \nu_{\rm eq} &= 441 \mbox{ Hz} 
        			\left(\frac{B_s}{10^9 {\rm~G}}\right)^{-6/7} 
        			\left(\frac{R_s}{10{\rm~km}} \right)^{-18/7} \nonumber \\
                  &\times 
                  	\left(\frac{\dot{M}}{\dot{M}_E} \right)^{3/7} 
					\left(\frac{M}{1.4~M_\odot} \right)^{5/7}
    \end{align}
%    \begin{eqnarray}
%    \nu _{eq} & = & 441 \mbox{~Hz} \left(\frac{B_s}{10^9 {\rm~G}}\right)^{-6/7} 
%    					\left(\frac{R_s}{10{\rm~km}} \right)^{-18/7} \nonumber \\
%              && \times \left(\frac{\dot{M}}{\dot{M}_E} \right)^{3/7} \left(\frac{M}{1.4~M_\odot} \right)^{5/7}
%    \end{eqnarray}
    where $\dot{M}_E=1.5\times 10^{-8}~M_\odot \mbox{ yr}^{-1}$ is the Eddington accretion rate.

\begin{itemize}
	\item In our analysis we assumed that channelled accretion onto the neutron 
	star can only take place when the disk is truncated inside the co-rotation
    radius. However, the large range of X-ray luminosities, and accordingly mass
    accretion rates, observed for AMXPs suggests that mass accretion onto the
    neutron star might persist even when the inner edge of the disk moves
    outside the co-rotation radius \citep[see e.g.][]{rappa04}, which indeed
    appears to be confirmed by observation \citep{bult15}.

    As pointed out by \citet{sprui93} the inner edge of the disk must have receded
    to $r_t \sim 1.3~r_c$ before the centrifugal force is strong enough to
    accelerate matter beyond the escape velocity and thus drive an outflow. If we
    consider the possibility that accretion may still occur for radii up-to
    $1.3~r_c$, then we find that the upper limit to the magnetic field strength
    increases by $58\%$ ($B_{\rm max} \propto (r_t/r_c)^{7/4}$) at most. The lower
    limit, being independent of the co-rotation radius, is unaffected.   
	\end{itemize}

\subsection{Observational sampling}
    \begin{itemize}
    \item 
    In order to determine the upper limit to the magnetic field strength we
    consider the lowest observed flux for which pulsations are significantly
    detected. However, as the flux decays, the signal-to-noise ratio also
    decreases, such that the non-detection of pulsations could be due to
    limited statistics. That is, the pulsations may persist below our detection
    limit. This concern is particularly relevant to XTE J1807.4--294, IGR~J17498--2921, and
    IGR~J17511--3057 in which pulsations are detected at approximately the same
    flux as the background level. If pulsations are still present at a very low
    level, and accretion is ongoing at even lower luminosities, then our estimates for
    these sources are overly conservative as we are overestimating the upper limit 
    to the magnetic field strength. Since $B_{\rm max} \propto L^{1/2}$, a future 
    detection of pulsations at a lower luminosity than reported in this work can 
    therefore be used to further constrain the range of magnetic field strengths.
    This is especially relevant for other X-ray satellites that have better 
    sensitivity and a lower background contamination such as XMM-Newton, or for
    a future ASTROSAT \citep{agrawal2006,singh14} or LOFT \citep{feroci12} mission.
  
    \item 
    For the lower limit to the magnetic field strength we use the highest
    observed flux for which pulsations are detected. As we note in
    Sec.~\ref{sec.method} the pulsations are expected to disappear at high flux
    when the disk extends to the neutron star surface, yet this is never
    observed. For all sources the highest observed flux considered always shows 
    pulsations, which implies that the inner edge of the disk never extends down 
    to the neutron star surface. Indeed some of the better sampled AMXPs (e.g.
    SAX~J1808.4--3658 and IGR~J00291+5934) show peak luminosities that vary by a
    factor of 2 between outbursts. If a future outburst reaches a higher peak
    luminosity than considered in this work, it will increase the lower limit
    as $B_{\rm min} \propto L^{1/2}$ and thus further constrain the allowed range 
    of magnetic field strength. 

    \end{itemize}

\subsection{Luminosity estimates}
    \begin{itemize}
    \item 
    To calculate the flux we consider the 3--20 keV X-ray band, which we convert
    to the bolometric flux by applying a correction factor \citep{gilfa98, 
    casel08, gallo02, campa03}. To be conservative in our estimates of the
    field strength we used a $\epsilon_{\rm bol} = 2$ for the upper limit and
    $\epsilon_{\rm bol} = 1$ for the lower limit.  However, if the correction
    factor is well constrained then this approach is overly pessimistic.
    For many accretion powered pulsars the correction factor tends to be within 
    $\sim10\%$ of 2 \citep{gallo08}, such that the error introduced in the magnetic 
    field estimate is only $\sim2\%$ ($B_{\rm max}\propto(\epsilon_{\rm bol}/2)^{1/2}$, 
    $B_{\rm min} \propto (\epsilon_{\rm bol}/1)^{1/2}$).
    If we adopt the same bolometric correction factor for the lower limit also,
    it we find the our estimates can improve by up-to $\lesssim30\%$.
    
    \item 
    Another source of uncertainty comes from the background contribution to the
    measured flux. We estimate the background contribution from the {\it RXTE}
    observations at the end of an outburst, when the source no longer shows 
    pulsations and has presumably returned to quiescence. If this estimate
    contains residual source emission, or the background contribution itself
    is variable, then this approach introduces an error in our field strength
    limits.
    
    Because the highest observed flux with pulsations is always much higher than 
    the background the lower limit to the magnetic field strength will not be 
    greatly affected by the uncertainty in the background estimate. The lowest
    observed flux with pulsations, however, is often comparable to the background
    contribution, so the effect on the upper limit needs to be considered carefully.
    
    For some sources the background flux is sufficiently lower than the minimum pulsating
    flux (as shown in Table~\ref{table.fluxstates}) that the effect of
    the background correction amounts to only a small change in the field estimates (e.g.
    $\sim 8\%$ for XTE~J1814--38, $\sim 12\%$ for Swift~J1756.9--2508).
    For sources where background flux could not be measured (HETE~J1900.1--2455 and
    XTE~J0929--314) the error introduced by neglecting the background is expected
    to similarly be only a few percent. 
    
    The remaining sources have a comparatively high background contribution, such that 
    the estimated background flux is a large fraction of the minimum pulsating flux. For 
    these AMXPs we conservatively took the lowest observed flux with pulsations as an upper
    limit to the real flux at which pulsations and thus accretion stops. We then calculated
    the upper limit on the magnetic field strength without adjusting for the background. 
    Accounting for the background would lower the source flux estimate by up-to a factor
    of two, and thus improve our estimates by roughly $\lesssim 40\%$. Further improvement
    might be achievable with more sensitive instrumentation (as noted in the previous section).

    \item 
    To convert the observed flux to the source luminosity we assumed 
    an isotropic emission process, however, the flux includes
    contributions from the hotspot, which may have a significant
    anisotropy \citep{poutanen06,viiro04,poutanen03}.
    
    The effect of anisotropic emission is not at all clear.
    The degree of anisotropy of the hotspot emission depends on
    assumptions of the emission process and can vary by a factor 
    of $\sim 2$ \citep{poutanen06}. Furthermore, what fraction of the 
    total flux is affected by this will depend on the size, shape and 
    position of the hotspot and is subject to considerable uncertainty.
    At best this effect applies only to the pulsed component of the 
    emission ($\sim10\%$) and thus introduces a systematic error
    in our estimates of only a few percent.
   	At worst most of the observed flux originates from a large hotspot that
	has a slightly beamed emission pattern. In that case the allowed magnetic 
	field range may show a systematic shift of up-to $\sim40\%$.

	\end{itemize}

\section{Comparison with previous works} \label{sec.summary}
In this work we estimated upper and lower limits to the magnetic field
strength of all accreting millisecond X-ray pulsars (AMXPs) observed with {\it RXTE}.
We assume that the detection of X-ray pulsations signifies ongoing magnetically channelled 
accretion. Thus, by associating the range of luminosity for which pulsations
are detected with the expected extent of the disk truncation radius, we have constrained
the magnetic dipole moment of the neutron star. The obtained equatorial 
surface magnetic field strengths of the 14 AMXPs analysed are presented 
in Table~\ref{table.fieldvalues}. 

Our magnetic field strength estimates are subject to a number of uncertainties,
which were discussed in the previous section. We note that we have
chosen most of the uncertain parameters such that we obtained a conservative
range for the magnetic moment. If parameters such as $\gamma_B$ or $\epsilon_{\rm bol}$
can be established more accurately, they will tighten the constraints further.
Errors in other parameters, such as the distance to the source, introduce a 
systematic shift in our results. Refinement in the measurement of these parameters 
would affect both the upper and the lower limits the same way.  Similarly, the uncertainties in the estimation of the background fluxes also affect both limits to the magnetic field. If better estimates are available from more sensitive instruments and also if there are future outbursts with flux ranges wider than those in the current work, the constraints on the magnetic field in Table~\ref{table.fieldvalues} can be easily updated by correcting the fluxes presented in Table~\ref{table.fluxstates} and recomputing the limits to the dipole moment using eq.~\ref{eq.mumin} and eq.~\ref{eq.mumax}.

For 5 of the 14 considered AMXPs (Swift~J1749.4--2807, IGR~J17511--3057, 
NGC~6440 X-2, XTE J1807.4--294 and IGR~J17498--2921) we obtain constraints
on the magnetic field strength for the first time.
For the other 9 AMXPs field strength estimates have been previously reported.
Below we discuss some of the techniques used to obtain those estimates and
how they compare with limits we report here. 

\begin{enumerate}
\item {\it Vacuum dipole radiation in quiescence:} 
For some AMXPs the spin frequency has been measured for successive outbursts
that are months or years apart. Measurements of spin-down during the intervening 
periods of quiescence, can then be used to estimate the magnetic field strength
by assuming the spin-down is due to magnetic dipole emission.
Such estimates have been obtained for IGR~J00291+5934 \citep{patru10c, hartm11}, 
XTE~J1751--305 \citep{riggi11b}, SAX~J1808.4--3658 \citep{hartm08,hartm09,patru12b} and  
Swift~J1756.9--2508 \citep{patru10c}. 
While the radio emission associated with magnetic braking has not yet been detected 
in any of these sources, the recent discovery of millisecond neutron stars pulsating 
alternately in X-rays and radio \citep{papit13b, archi14, papit15} provides some 
evidence that this interpretation of measured spin-down is correct.

Our magnetic field strength estimates appear to be systematically lower
than those obtained through quiescent spin down, although we note that
given the systematic uncertainties discussed in the previous section, the
results of these two approaches are roughly consistent. 

\item{\it Quiescent luminosity estimates:}
For some LMXBs e.g. SAX~J1808.4--3658, Aql X-1 \citep{salvo03}, KS~1731--260
\citep{burde02} and XTE~J0929--314 \citep{wijna05}, limits to the magnetic
field strength have been inferred from measurements of the quiescent X-ray
luminosity.  
However, given the very low count rates in the quiescent phase, and a poor understanding
of which physical mechanism governs the radiation process, these methods offer less reliable constraints 
on the dipole moment compared to other approaches. 
While for Aql X-1 no other independent confirmation of the upper limits exists, the upper 
limits for SAX~J1808.4--3658 and XTE~J0929--314 obtained through quiescent luminosity methods
are comparable to the upper limits we report here.

\item{\it Accretion induced spin down estimates:} 
For some systems the magnetic field strength has been estimated by comparing
observed rate of spin down during ongoing accretion to theoretical estimates of
magnetic torque.  For example, for XTE~J1814--338 \citet{papit07} assume that 
observed pulse frequency variations are caused by spin down due to the
torque applied by an accretion disk that is truncated near the
co-rotation radius. Following theoretical calculations of accretion induced
spin down torques \citep{rappa04}, the authors estimate the surface magnetic field to be
$\sim 8\times 10^8$ G, which is comparable our conservative upper limit to the field, and
significantly higher than the upper limit we obtain using eq. \ref{eq.newgamma}. 
However, given their simplifying assumptions regarding the magnetic field topology at the Alfv\'en 
radius, and the considerable uncertainty in interpreting pulse frequency variations
as spin variations \citep{patru09b}, accretion induced spin down estimates are less robust than
those we obtain. 

\item{\it Burst oscillations:}
In some accreting pulsars the phase of burst oscillations in type I X-ray
bursts is locked to the phase of the accretion powered pulsations (XTE~J1814--338 
\citealt{watts08}, IGR~J17480--2446 \citealt{cavec11}). \citet{cavec11} argue that
if this phase-locking is due to magnetic confinement of the flame propagation front, then
it would require a field strength of $\gtrsim 5\E9$~G. Even given the systematic
uncertainties that enter in our estimates, such a large magnetic field strength
would be difficult to reconcile with the lower upper limit we obtain
for XTE~J1814--338.

\item{\it Spectral state transitions in LMXBs:} 
In another approach, spectral state transitions have been used to identify 
the onset of the propeller regime \citep{matsu13,asai13}. These authors argue that a
spectral change and a fast decline of luminosity towards the end of an outburst indicates the onset of 
the propeller regime and thus an accretion disk that is 
truncated at the co-rotation radius. 
Such spectral state transitions have been investigated for only a handful of LMXBs (e.g. Aql
X-1, 4U~1608--52, XTE~J1701--462), but give limits on the magnetic field
strength of Aql X-1 that are tighter than those we obtained.
However, there is no clear evidence that these transitions are indeed caused by a
propeller effect. In fact, the observation of similar state transitions in black hole 
binaries seems to suggest otherwise \citep{jonke04a}.

\end{enumerate}

To conclude, the magnetic field estimates we obtain agree with
most of the other indirect methods to within an order of magnitude. The large
uncertainties on many parameters, as well as the uncertainty in the underlying
assumptions, introduce a significant spread in the range of field strengths 
inferred via each approach, with no single technique being more robust than
the others. Nonetheless, all alternative methods discussed in this section
require that the AMXP be observed during a specific state of its outburst (e.g
during quiescence or a spectral state transition). Since observations of such
special states are not available, or may not even exist, for all sources, these
methods are not suitable for studying the population. By contrast, the method
used in this work applies to any AMXP, making it more reliable for comparing
the field strengths of the population and understanding the evolutionary
processes which lead to the formation of AMXPs.

\section{Acknowledgements}
We thank Deepto Chakrabarty, Dimitros Psaltis, Alessandro Patruno, Anna Watts and Sushan
Konar for constructive suggestions and helpful discussions which improved the
content and presentation if this work. We thank Tolga G{\"u}ver for his careful scrutiny
 and comments which helped improve the manuscript. DM thanks CSIR India for the junior
research fellow grant 09/545(0034)/2009-EMR-I. DM also acknowledges the
hospitality of Anton Pannekoek Institute where part of the work was carried out
during his visit. PB and MK acknowledge support from the Netherlands Organisation 
for Scientific Research (NWO).
The work has made use of the public archival data, part of
which is taken from observations originally proposed by the following PIs:
Duncan Galloway, Diego Altamirano, Jean Swank, Rudy Wijnands and Wei Cui; the
rest being Target of opportunity (TOO) observations of source outbursts.

%\appendix
%\input{timing.tex}
%\input{spectra.tex}

%==========================================
%----------------- Bibliography and bibfile
\def\apj{ApJ}%
\def\mnras{MNRAS}%
\def\aap{A\&A}%
\def\apjl{ApJ}
\def\physrep{PhR}
\def\apjs{ApJS}
\def\pasa{PASA}
\def\pasj{PASJ}
\def\nat{Nature}
\def\memsai{MmSAI}
\def\aj{AJ}%
\def\aaps{A\&AS}%
\def\iaucirc{IAU~Circ.}%
\def\sovast{Soviet~Ast.}%
\def\apss{Ap\&SS}

\bibliographystyle{mn2e}
\bibliography{dip_1.5.15}

\clearpage
\appendix
% !TEX root =  amxp.tex

\section{Details of the timing analysis}
\label{sec.timingdetails}
    \subsection{Swift J1756.9--2508:}
        The two outbursts of Swift J1756.9--2508 were observed with {\it RXTE} under
        program IDs P92050 \& P93065 (2007) and P94065 (2009). We barycenter this
        data using the best known coordinates of \citet{krimm07b}. For the
        timing solution we adopt the ephemeris of \citet{patru10b}.
        
    \subsection{XTE J0929--314:}
        We correct the {\it RXTE} data (program ID P70096) to the Solar system barycenter
        using the Chandra source position of \citet{juett03}. The data was folded 
        using the timing solution of \citet{iaco09}. 
    
    \subsection{XTE J1807.4--294:}
        The {\it RXTE} data of the outburst of XTE J1807.4--294 is given by program IDs P70134 
        and P80(145/419). We use the Chandra coordinates of \citet{markw03b} to barycenter
        the data. We use the system ephemeris of \citet{riggi08} to fold the data. 

    \subsection{NGC 6440 X-2:}
        The nine short outbursts of NGC 6440 X-2 that were observed with {\it RXTE}
        are given by program IDs P94044, P94315 and P95040 \citep{patru13b}. We correct the data
        to the Solar System barycenter using the Chandra position of \citet{heink10}.
        The timing solution for the first outburst is given by \citet{altam10c}. This 
        timing solution is also used the later outbursts, but with the locally optimized
        time of ascending node values given in Table~\ref{tab.ngc6440tasc}.
        \begin{table}
            \centering
            \caption{Adopted $T_{\rm asc}$ values for the outbursts of NGC 6440 X-2.}
            \label{tab.ngc6440tasc}
            \begin{tabular}{|l l|}
            \hline	 
            \multicolumn{2}{c|}{\bf NGC 6440 X-2} \\
            Outburst & $T_{\rm asc}$ \\
            \hline
                Outburst 1   & 55042.817   \\
                Outburst 2   & 55073.034   \\
                Outburst 3   & 55106.012   \\
                Outburst 4   & 55132.907   \\
                Outburst 5   & 55276.625  \\
                Outburst 6   & 55359.470  \\
                Outburst 7   & 55473.854   \\
                Outburst 8   & 55584.714   \\
                Outburst 9   & 55871.231   \\
            \hline
            \end{tabular}
        \end{table} 

    \subsection{IGR~J17511--3057:}
       The {\it RXTE} data of the 2009 outburst of IGR~J17511--3057 is given by program ID
       P940(41/42). We barycenter the data using the Chandra position given by
       \citet{nowak09}. The timing solution of this source is given by \citet{riggi11a}.

    \subsection{XTE J1814--338:}
        The {\it RXTE} data of XTE J1814--338 is given by program IDs P80(138/145/418) and P92054.
        The optical position used to barycenter the data is given by \citet{krauss05} and
        the timing solution is taken from \citet{papit07}.

    \subsection{HETE J1900.1--2455:}
        The source HETE J1900.1--2455 has a long history of activity and thus its {\it RXTE}
        data is spread over many program IDs: P910(15/57/59), P91432, P92049, P93(030/451), 
        P940(28/30), P95030 and P96030. For source coordinates we use the optical position
        of \citet{fox05}. The timing solution is given by \citet{patru12}.

    \subsection{SAX J1808.4--3658:}
        The {\it RXTE} program IDs for SAX J1808.4--3658 are given by P30411 (1998), P40035 (2000),
        P70080 (2002), P91056 \& P91418 (2005), P93027 \& P93417 (2008) and P96027 (2011).
        When considering the outbursts of SAX J1808.4--3658 we exclude the prolonged outburst
        tail which represents an unusual disk state \citep{patru09d,patru15}, so the outburst of 2000 is 
        entirely omitted from the analysis. Source coordinates
        for barycentering are taken from \citet{hartm08}. For the ephemeris we adopt
        the solutions of \citet{hartm08, hartm09} and \citet{patru12b} for the respective
        outbursts of 1998, 2002 and 2005; 2008; and 2011.

    \subsection{IGR~J17498--2921:}
        {\it RXTE} data for IGR~J17498--2921 is given by program ID P96435. For the source position 
        we adopt the Chandra coordinates of \citet{chakra11}. The timing solution is given
        by \citet{papit11b}.

    \subsection{XTE J1751--305:}
        We only analyze the {\it RXTE} data of the 2002 outburst (see main text for details) which
        is given by programs P70131 and P70134. To barycenter the data we use the Chandra position
        of \citet{Markwardt2002b}. For the orbital ephemeris we use the timing solution of \citet{Markwardt2002c}.

    \subsection{SAX J1748.9--2021:}
        The {\it RXTE} data of SAX J1748.9--2021 is given by programs P30425, P60035 \& 
        P60084 (2001), P91050 (2005) and P94315 (2010). We use the Chandra position
        of \citet{intza01} for barycentering the data. The timing solution of
        \citet{patru09e} was used to fold the data of the 1998, 2001 and 2005 outbursts
        and that of \citet{patru10a} for the 2010 outburst.
        
%\begin{table}
%    \centering
%    \caption{Timing model for the 2010 outburst of Swift J1749.4--2807.}
%    \label{tab.timingpar1}
%    \begin{tabular}{|l  l|}
%    \hline	 \multicolumn{2}{c|}{\bf Swift J1749.4--2807} \\
%    Parameters 	& Values \\
%    \hline  
%        Orbital period, $P_{\rm orb}$ (days)             & 0.365219(1) \\
%        Projected semimajor axis, $a_x\sin(i)$ (lt-s)   & 0.38760(4)  \\
%        Time of ascending node, $T_{asc}$ (MJD)         & 52191.87505(1)  \\
%        Spin frequency, $\nu_0$ (Hz)                    & 442.36108118(5) \\
%    \hline
%    \end{tabular}
%\end{table} 

    \subsection{Swift J1749.4--2807:}
        We use the data of {\it RXTE} program P95085. For barycentering we adopt the
        X-ray coordinates of \citet{wijna09}, obtained with XMM-Newton. We use
        the ephemeris of \citet{altam11a} to fold the data. 

    \subsection{IGR~J00291+5934:}
        We use {\it RXTE} data from programs P90052 and P90425 for the 2004 outburst and
        program P93013 and P93435 for the 2008 outburst. We use radio coordinates
        of \citet{rupen04} to barycenter the data. The timing solution of \citet{patru10c}
        was used to fold the data. 

%   \subsection{Aql X-1:}
%       Not applicable.

% !TEX root =  amxp.tex
\section{Details of the spectral analysis} \label{sec.fitdetails}
For the spectral analysis we extracted the data of all Xenon layers of PCU2, and
for observations with a poor signal to noise ratio we combined data from all active
PCUs.
Due to poor signal to noise some spectral fits at low flux have a
reduced-$\chi^2$ of much less than 1. As our work does not focus on obtaining
the most accurate spectral model, but rather on measuring the flux, our final
results will not be significantly affected by the model uncertainty of these
fits.

Our analysis of the {\it RXTE/PCA} data was performed for the $3-20$~keV energy range
using {\scshape XSPEC} version 12.7.1.  The galactic absorption was modelled with the
{\it TBABS} model \citep{wilms00}. Since {\it RXTE} instruments
cannot properly constrain the galactic absorption due to neutral hydrogen in
the lower energy range, we fix $N_H$ to values obtained from literature. The errors quoted
are for a $95\%$ confidence limit.

\subsection{Swift J1756.9--2508:}

\begin{table*}
\centering
\caption[%
    Spectral fit parameters of Swift J1756.9--2508.
]{\small%
    Spectral fit parameters of Swift J1756.9--2508.
}
\label{tab.J1756}
\begin{tabular}{|l  | l  l  l |}
\hline
 & \multicolumn{3}{c|}{\bf Swift J1756.9--2508} \\
 & High flux & Low flux & Background \\
\hline 
ObsIDs                          & 94065-02-01-05			& 94065-06-02-03			& 94065-06-03-02  \\[1.5mm]
$N_H (10^{22}\mbox{ cm}^{-2})$	& 5.4 (fixed)				& 5.4 (fixed)				& 5.4 (fixed)     \\[1.5mm]
$T_{\rm BB}$	(keV)           & --       		            & $0.49 \pm 0.08$		& --			  \\[1.5mm]
${\rm Norm}_{\rm bb}$           & --                        & $10^{+6}_{-2} \E{-4}$    	& --			  \\[1.5mm]
$\Gamma$			            & $1.970 \pm 0.014$	        & $2.24 \pm 0.06$		& $2.67 \pm 0.11$ \\[1.5mm]
$\chi ^2$/d.o.f.		        & 45.66/37				    &  40.39/41				& 34.68/36		  \\[1.5mm]
Flux ($\ergcms$)                & $6.30 \pm 0.04 \E{-10}$   & $1.990 \pm 0.017 \E{-10}$   & $4.07 \pm 0.15 \E{-11}$    \\
\hline
\end{tabular} \\ \medskip
\flushleft
	$N_H$ is the column density of neutral hydrogen for the {\it tbabs} model, 
    $T_{\rm bb}$ is the black body temperature, 
    ${\rm Norm}_{\rm bb}$ is the normalisation of the {\it bbody} model, 
    $\Gamma$ is the power-law index.
%    $E_{\rm cut}$ is the high energy cutoff for a cutoff power-law model, 
%    $T_{\rm disk}$ is the temperature of the inner radius of the accretion disk, 
%    and ${\rm Norm}_{\rm disk}$ is the normalisation of the {\it diskbb} model. 
    The first column represents the highest flux (HF) with pulsations, the second 
    column gives the lowest flux (LF) with pulsations and the third column gives
    the observation used to measure the background emission (BE).
\end{table*}

For ObsIDs 94065-02-01-05 (high flux with pulsations, HF) and 94065-06-02-03 (low flux with pulsations, LF) data were extracted
from PCU2 whereas for 94065-06-03-02 (background estimate, BE), data from PCU 0, 2 and 4 were
combined for the spectral analysis. The HF and BE spectra were fit with a
simple absorbed power-law model {\it XSPEC} \citep{arnau96}. The LF spectrum was fit with a power-law
continuum and a thermal black body component. The  hydrogen column density accounting for galactic absorption  was
fixed to $N_H = 5.4 \times 10^{20} \mbox{ cm}^{-2}$ \citep{krimm07b}. A Gaussian emission
component centred at $\sim 6.5$~keV (width fixed to $10^{-3}$~keV) was required to obtain statistically acceptable
fits. The results of the spectral analysis are presented in Table~\ref{tab.J1756}.

\subsection{XTE J0929--314:}

\begin{table*}
\centering
\caption[%
    Spectral fit parameters of XTE J0929--314.
]{\small%
    Spectral fit parameters of XTE J0929--314, see Table~\ref{tab.J1756}
    for details.
}
\label{tab.J0929}
\begin{tabular}{|l  | l  l  l |}
\hline
 & \multicolumn{3}{c|}{\bf XTE J0929--314} \\
 & High flux & Low flux & Background \\
\hline 
ObsIDs                          & 70096-03-02-00 			& 70096-03-14-00	 		& -- \\[1.5mm]
$N_H  (10^{22}\mbox{ cm}^{-2})$	& 0.0076 (fixed) 			& 0.0076 (fixed)		    & -- \\[1.5mm]
$T_{\rm BB}$	(keV)			& $0.82 \pm 0.04$			& $0.53 \pm 0.05$			& -- \\[1.5mm]
${\rm Norm}_{\rm bb}$			& $3.5  \pm 0.5 \E{-3}$		& $1.9 \pm 0.4 \E{-3}$		& -- \\[1.5mm]
$\Gamma$			        	& $1.80 \pm 0.04$			& $1.90  \pm 0.11$			& -- \\[1.5mm]
$\chi ^2$/d.o.f.				& 36.33/42					& 26.11/42					& -- \\[1.5mm]
Flux ($\ergcms$)                & $4.42 \pm 0.02 \E{-10}$   & $6.64 \pm 0.14 \E{-11}$   & -- \\
\hline
\end{tabular}
\end{table*}

For ObsID 70096-03-02-00 (HF) data were extracted from PCUs 0, 2, 3 \& 4 whereas
for 70096-03-14-00 (LF) they were extracted from PCUs 0, 2, 4. The spectra were 
fit with a power-law continuum and thermal blackbody component. The hydrogen 
column density was fixed to $N_H = 7.6\E{20} \mbox{ cm}^{-2}$ \citep{juett03}. 
The results are presented in Table~\ref{tab.J0929}.

\subsection{\bf XTE J1807.4--294:}

\begin{table*}
\centering
\caption[%
    Spectral fit parameters of XTE J1807.4--294.
]{\small%
    Spectral fit parameters of XTE J1807.4--294, see Table~\ref{tab.J1756}
    for details.
}
\label{tab.J1807}
\begin{tabular}{|l  | l  l  l |}
\hline
 & \multicolumn{3}{c|}{\bf XTE J1807.4--294} \\
 & High flux & Low flux & Background \\
\hline 
ObsIDs                          & 70134-09-02-01			& 80145-01-17-02			& 80419-01-01-01    \\[1.5mm]
$N_H (10^{22}\mbox{ cm}^{-2})$ 	& 0.56 (fixed)			    & 0.56 (fixed)				& 0.56 (fixed) 	    \\[1.5mm]
$\Gamma$		        		& $1.90 \pm 0.03$			& $2.19 \pm 0.04$			& $2.16 \pm 0.04$   \\[1.5mm]
$T_{\rm BB}$	(keV)			& $1.46^{+0.33}_{-0.19}$	& --						& -- 				\\[1.5mm]
${\rm Norm}_{\rm bb}$			& $4.9 \pm 0.2 \E{-4}$		& --						& --				\\[1.5mm]
$\chi ^2$/d.o.f.	        	& 49.87/42					& 31.07/44					& 34.69/44		    \\[1.5mm]
Flux ($\ergcms$)                & $8.19 \pm 0.04 \E{-10}$   & $3.51 \pm 0.07 \E{-10}$   & $7.25 \pm 0.15 \E{-11}$    \\
\hline
\end{tabular}
\end{table*}

For ObsIDs 70134-09-02-01 (HF) and 80145-01-17-02 (LF) the data were
extracted from PCU2 whereas for 80419-01-01-01 (BE) data were extracted from
PCUs 0, 2 and 3. The HF spectrum was fit with a power-law continuum and thermal
blackbody component whereas the LF and BE spectra were fit with a simple power-law
model. The hydrogen column density was fixed to $N_H = 5.6 \times 10^{21}
\mbox{ cm}^{-2}$ \citep{falan05}. The resulting fit parameters are presented 
in Table~\ref{tab.J1807}.

\subsection{NGC 6440 X-2:}

\begin{table*}
\centering
\caption[%
    Spectral fit parameters of NGC 6440 X-2.
]{\small%
    Spectral fit parameters of NGC 6440 X-2, see Table~\ref{tab.J1756}
    for details.
}
\label{tab.ngc6440}
\begin{tabular}{|l  | l  l  l |}
\hline
 & \multicolumn{3}{c|}{\bf NGC 6440 X-2} \\
 & High flux & Low flux & Background \\
\hline 
ObsIDs                          & 94044-04-02-00			& 96326-01-40-01			& 96326-01-36-00 \\[1.5mm]
$N_H  (10^{22}\mbox{ cm}^{-2})$ & 0.59 (fixed)				& 0.59 (fixed)				& 0.59 (fixed)	 \\[1.5mm]
$\Gamma$		       			& $1.83 \pm 0.02$			& $2.36^{+0.21}_{-0.12}$	& $2.3 \pm 0.4$ \\[1.5mm]
$\Gamma$ (Cutoffpl)				& --						& $1.1 \pm 0.8$				& --			\\[1.5mm]
$E_{\rm cut}$					& --						& $5.5^{+9}_{-2}$			& --			\\[1.5mm]
$\chi ^2$/d.o.f.	    		& 34.38/38					& 45/35						& 36.77/44		 \\[1.5mm]
Flux ($\ergcms$)                & $2.62 \pm 0.03 \E{-10}$   & $3.4 \pm 0.2 \E{-11}$    	& $1.3\pm0.2\E{-11}$  \\
\hline
\end{tabular} \\ \medskip
\flushleft
    $E_{\rm cut}$ is the high energy cutoff for a cutoff power-law model.
\end{table*}

For ObsIDs 94044-04-02-00 and 96326-01-36-00 data was extracted from PCU2 whereas 
for ObsID 96326-01-40-01 data was taken from PCU 1 and 2. The hydrogen column density 
was fixed to $N_H = 5.9 \times 10^{21} \mbox{ cm}^{-2}$ \citep{harri96}. The data were 
fit with a simple absorbed power-law for all ObsIDs. The fit parameters are presented 
in Table~\ref{tab.ngc6440}.

\subsection{IGR~J17511--3057:}

\begin{table*}
\centering
\caption[%
    Spectral fit parameters of IGR~J17511--3057.
]{\small%
    Spectral fit parameters of IGR~J17511--3057, see Table~\ref{tab.J1756}
    for details.
}
\begin{tabular}{|l  | l  l  l |}
\hline
 & \multicolumn{3}{c|}{\bf IGR~J17511--3057} \\
 & High flux & Low flux & Background \\
\hline 
ObsIDs                              & 94041-01-01-02			 & 94042-01-03-04		  & 94042-01-02-05      \\[1.5mm]
$N_H\times 10^{22}\mbox{ cm}^{-2}$  & 1 (fixed)	 		         & 1 (fixed)			  & 1 (fixed)			\\[1.5mm]
$T_{\rm BB}$	(keV)	            & $1.10 \pm 0.16$	         & --          		  	  & -- 				    \\[1.5mm]
${\rm Norm}_{\rm bb}$		    	& $6 \pm 3 \E{-4}$   		 & --			  	  		  & --				    \\[1.5mm]
$\Gamma$			    			& $1.70 \pm 0.04$			 & $2.02 \pm 0.04$		  		& $2.37 \pm 0.06$	    \\[1.5mm]
$\chi ^2$/d.o.f.		 	 	  	& 44.81/39					 & 39.89/41	                	& 36.71/44			\\[1.5mm]
Flux ($\ergcms$)                    & $8.65 \pm 0.05 \E{-10}$    & $1.000 \pm 0.018 \E{-10}$ 	& $6.960 \pm 0.018 \E{-11}$      \\
\hline
\end{tabular}
\label{tab.J17511}
\end{table*}

For ObsID 94041-01-01-02 (HF) the data were extracted from PCU 2.
For ObsIDs 94042-01-03-04 (LF) and 94042-01-02-05 (BE) spectral analysis was
performed by combining data from PCUs 2, 3 and PCUs 2, 4 respectively. The
hydrogen column density was fixed to $N_H = 1\times 10^{22} \mbox{ cm}^{-2}$
\citep{papit10, paizi12}. The HF was well described with a power-law continuum 
and thermal blackbody component, whereas the LF and BE spectra were
fit with a simple absorbed power law. A Gaussian feature at $\sim 6.5$~keV was
added to the HF and BE spectra to improve the fits. The resulting parameters are
presented in Table~\ref{tab.J17511}.

\subsection{XTE J1814--338:}

\begin{table*}
\caption[%
    Spectral fit parameters of XTE J1814--338.
]{\small%
    Spectral fit parameters of XTE J1814--338, see Table~\ref{tab.J1756}
    for details.
}
\label{tab.J1814}
\centering
\begin{tabular}{|l  | l  l  l |}
\hline
 & \multicolumn{3}{c|}{\bf XTE J1814--338} \\
 & High flux & Low flux & Background \\
\hline 
ObsIDs                          & 80418-01-03-08			& 80418-01-07-08			& 80418-01-09-00    \\[1.5mm]
$N_H (10^{22}\mbox{ cm}^{-2})$	& 0.167 (fixed) 			& 0.167 (fixed)				& 0.167 (fixed)		\\[1.5mm]
$T_{\rm BB}$	(keV)			& $1.21 \pm +0.06$			& --						& --   			\\[1.5mm]
${\rm Norm}_{\rm bb}$			& $1.7 \pm 0.4 \E{-3}$		& --						& --			    \\[1.5mm]
$\Gamma$						& $1.55  \pm 0.04$			& $1.96 \pm 0.03$			& $2.4 \pm 0.3$	    \\[1.5mm]
$\Delta \chi ^2$/d.o.f.			& 63.99/42					& 23.2/43					& 28.17/44		\\[1.5mm]
Flux ($\ergcms$)                & $4.41 \pm 0.03 \E{-10}$   & $6.00 \pm 0.10 \E{-11}$   & $1.00 \pm 0.16 \E{-11}$    \\
\hline
\end{tabular}
\end{table*}

Since the observed count rate of XTE~J1814--338 is small 
($< 40 \mbox{ct~s}^{-1}$ for the entire outburst), data from all available PCUs 
were combined for all spectra. For ObsID 80418-01-03-08 (HF) the data were extracted from
PCU 0, 2 and 3. For 80418-01-07-08 (LF) data from PCUs
0, 1, 2 and 3 and for 80418-01-09-00 (BE) PCUs 0, 1 and 2 were used. The
hydrogen column density was fixed to $N_H = 1.67 \times 10^{21}~\mbox{cm}^{-2}$ 
\citep{kraus05}.  

The HF spectrum was fit with a power-law continuum and a blackbody component, 
whereas the LF and BE spectra were fit with a simple absorbed power-law model. 
The details of the fit parameters are presented in Table~\ref{tab.J1814}.

\subsection{HETE J1900.1--2455:}

\begin{table*}
\centering
\caption[%
    Spectral fit parameters of HETE J1900.1--2455.
]{\small%
    Spectral fit parameters of HETE J1900.1--2455, see text and Table~\ref{tab.J1756} for further details.
}
\label{tab.hete}
\begin{tabular}{|l  | l  l  l |}
\hline
 & \multicolumn{3}{c|}{\bf HETE J1900.1--2455} \\
 & High flux & Low flux & Background \\
\hline 
ObsIDs                         	& 91015-01-06-00			& 91059-03-02-00		  	& --  \\[1.5mm]
$N_H (10^{22}\mbox{ cm}^{-2})$ 	& 0.1 (fixed)			    & 0.1 (fixed)				& --  \\[1.5mm]
$T_{\rm BB}$	(keV)		   	& 				    		& $0.74 \pm 0.10$			& --  \\[1.5mm]
${\rm Norm}_{\rm bb}$		   	&  				    		& $8    \pm 2 \E{-4}$ 		& --  \\[1.5mm]
$\Gamma$ (cutoffpl)		   		& $0.9 \pm 0.1$        	    & 							& --  \\[1.5mm]
$E_{\rm cut}$			   		& $4.0 \pm 0.2$ 	        &							& --  \\[1.5mm]
$\Gamma$		 	   			&                    	    & $1.83 \pm 0.07$ 	 	  	& --  \\[1.5mm]
$\chi ^2$/d.o.f.		   		& 42.74/40				    & 25.26/40					& --  \\[1.5mm]
Flux ($\ergcms$)                & $1.15 \pm 0.05 \E{-9}$   	& $3.84 \pm 0.03 \E{-10}$ 	& --  \\
\hline
\end{tabular}
\end{table*}

The data were extracted from all layers of PCU2 for the considered ObsIDs. The
hydrogen column density was fixed to $N_H = 1\times 10^{21} \mbox{ cm}^{-2}$
\citep{papit13}. We obtain the HF spectrum from ObsID 91015-01-06-00 with fit with
{\it tbabs*cutoffpl}. We found a large excess at $\sim 6$~keV, which was
modelled with a Gaussian centred at $\sim 6.2$~keV with a width of $\sim 1$~keV.
The resulting fits were statistically acceptable.  

The ObsID 91059-03-02-00 (LF) was fit with {\it tbabs(bbody+powerlaw)}. 
A weak Gaussian feature at $\sim 6.5$~keV was added
to improve the fits. The results are presented in Table~\ref{tab.hete}.

A background estimate is not available for this source.

\subsection{SAX J1808.4--3658:}

\begin{table*}
\centering
\caption[%
    Spectral fit parameters of SAX J1808.4--3658.
]{\small%
    Spectral fit parameters of SAX J1808.4--3658.
}
\label{tab.J1808}
\begin{tabular}{|l  | l  l  l |}
\hline
 & \multicolumn{3}{c|}{\bf SAX J1808.4--3658} \\
 & High flux & Low flux & Background \\
\hline 
ObsIDs                         	& 70080-01-01-04				& 30411-01-11-00		& 30411-01-11-02           \\[1.5mm]
$N_H (10^{22}\mbox{ cm}^{-2})$ 	& 0.2 (fixed) 				& 0.2 (fixed)				& 0.2 (fixed)              \\[1.5mm]
$\Gamma$ (Powerlaw)		   		& --						& $2.24 \pm 0.05$ 			& $2.23 \pm 0.11$ 	\\[1.5mm]
$T_{\rm bb}$	(keV)		   	& $1.25 \pm 0.07$			& --						& --          		\\[1.5mm]
${\rm Norm}_{\rm bb}$		   	& $9.2 \pm 1.4 \E{-3}$		& --						& --                       \\[1.5mm]
$T_{\rm disk}$	(keV)		   	& $0.45 \pm 0.06$			& --						& -- 	                   \\[1.5mm]
${\rm Norm}_{\rm disk}$		   	& $8.7^{+18}_{-5} \E{3}$	& --						& --	                   \\[1.5mm]	
$E_1$ 		(keV)		   		& $6.2$ (fixed)				& --						& --                       \\[1.5mm]
$\sigma _1$ 	(keV)		   	& $1.04 \pm 0.15 $	 		& --						& --                       \\[1.5mm]	
$\Gamma$ (NTHCOMP)		   		& $2.3 \pm 0.1$				& --						& --			   \\[1.5mm]
$T_e$           (keV)		   	& 100 (fixed)				& --						& --	   		   \\[1.5mm]
$\chi ^2$/d.o.f.		   		& 33.37/35					& 50.18/45					& 35.47/43                 \\
Flux ($\ergcms$)               	& $1.850 \pm 0.007 \E{-9}$  & $2.82 \pm 0.07 \E{-11}$ 	& $1.21 \pm 0.08 \E{-11}$  \\
\hline
\end{tabular} \\ \medskip
\flushleft
    $T_{\rm disk}$ is the temperature of the inner radius of the accretion disk, 
    and ${\rm Norm}_{\rm disk}$ is the normalisation of the {\it diskbb} model. 
	The fluorescent iron line is fit with a Gaussian component, such that $E_1$ is the line energy 
    and $\sigma _1$ the width of the line. Similarly $E_2$ and $\sigma _2$ are
    the line energy and width of a Gaussian component needed to model an excess 
    at $\sim 5.4$~keV. See text and Table~\ref{tab.J1756} for further details.
    
\end{table*}

For ObsID 70080-01-01-04 (HF) the data were extracted from the top layers of
PCU2. The hydrogen column density was fixed to $N_H = 2\E{21} \mbox{ cm}^{-2}$ \citep{ 
papit09, cacke09, patru09a}. The continuum was modelled with the thermal comptonisation
{\it NTHCOMP} of \citet{zdzia96, zycki99}. A blackbody and diskblackbody component were
also required to model the continuum. A broad excess at $\sim 6.2$~keV was seen in the 
residuals, reminiscent of a relativistically broadened iron line reported in other works 
\citep{papit09, patru09a, cacke09}. However, owing to the poor spectral resolution of RXTE, 
we have not employed the sophisticated relativistic models like {\it diskline} and instead 
modelled the feature with a broadened Gaussian. The central energy of the Gaussian was fixed 
to $6.2$~keV. 

Data for ObsIDs 30411-01-11-00 (LF) and 30411-01-11-02 (BE) was extracted from all PCUs and 
combined for spectral analysis. The data fit well with a simple power-law model. The results 
are presented in Table~\ref{tab.J1808}.

\subsection{IGR~J17498--2921:}

\begin{table*}
\centering
\caption[%
    Spectral fit parameters of IGR~J17498--2921.
]{\small%
    Spectral fit parameters of IGR~J17498--2921, see Table~\ref{tab.J1756}
    for details.
}
\label{tab.J17498}
\begin{tabular}{|l  | l  l  l |}
\hline
 & \multicolumn{3}{c|}{\bf IGR~J17498--2921} \\
 & High flux & Low flux & Background \\
\hline 
ObsIDs                         	& 96435-01-02-01			& 96435-01-06-04		  & 96435-01-07-01          \\[1.5mm]
$N_H (10^{22}\mbox{ cm}^{-2})$ 	& 2.87 (fixed) 				& 2.87 (fixed)			  & 2.87 (fixed)	        \\[1.5mm]
$T_{\rm BB}$	(keV)		   	& $1.51 \pm 0.05$			& $1.93 \pm 0.15$	  			& $1.76 \pm 0.08$	        \\[1.5mm]
${\rm Norm}_{\rm bb}$		   	& $1.9 \pm 0.2 \E{-3}$	    & $1.41^{+0.2}_{-0.08} \E{-3}$	& $2.2 \pm 0.2 \E{-3}$  \\[1.5mm]
$\Gamma$			   			& $1.93\pm 0.02$			& $2.39^{+0.1}_{-0.06}$		  	& $2.3 \pm 0.1$	        \\[1.5mm]
$\chi ^2$/d.o.f.		   		& 52.76/39						& 28.90/38			  		& 32.39/38		        \\[1.5mm]
Flux ($\ergcms$)               	& $1.130^{+0.003}_{-0.006} \E{-9}$  & $4.44 \pm 0.04 \E{-10}$ 	  & $4.23 \pm 0.04 \E{-10}$ \\
\hline
\end{tabular}
\end{table*}

Data for all ObsIDs viz. 96435-01-02-01 (HF),  96435-01-06-04 (LF) and 96435-01-07-01 (BE)
were extracted from all layers of PCU 2. All three spectra were fit with a simple power-law
continuum model and a thermal blackbody component. A narrow Gaussian component
centred at $\sim 6.5$~keV (most likely a feature from background emission) was
required to obtain a statistically acceptable fit. The hydrogen column density was
fixed to $N_H = 2.87\times 10^{22} \mbox{ cm}^{-2}$ \citep{torre11}. The results 
are presented in Table~\ref{tab.J17498}. 

We note that flux measured from the LF and BE observations are the same within error,
indicating that the lowest flux with pulsations is background dominated.

\subsection{XTE J1751--305:}

\begin{table*}
\centering
\caption[%
    Spectral fit parameters of XTE J1751--305.
]{\small%
    Spectral fit parameters of XTE J1751--305, see Table~\ref{tab.J1756}
    for details.
}
\label{tab.J1751}
\begin{tabular}{|l  | l  l  l |}
\hline
 & \multicolumn{3}{c|}{\bf XTE J1751--305} \\
 & High flux & Low flux & Background \\
\hline 
ObsIDs                         & 70134-03-01-00			& 70131-01-09-000			& 70131-02-04-00        \\[1.5mm]
$N_H (10^{22}\mbox{ cm}^{-2})$ & 1 (fixed) 				& 1 (fixed)					& 1 (fixed)			  \\[1.5mm]
$T_{\rm bb}$	(keV)		   & $1.95 \pm 0.16$		& --		  				& --		      \\[1.5mm]
${\rm Norm}_{\rm bb}$		   & $2.1  \pm 0.3 \E{-3}$	& --  						& --			 \\[1.5mm]
$T_{\rm disk}$	(keV)		   & --						& --		  				& --			      \\[1.5mm]
${\rm Norm}_{\rm disk}$		   & --						& --	  	  				& --				      \\[1.5mm]	
$\Gamma$			  		   & $1.77 \pm 0.03$			 & $2.070 \pm 0.007$	& $2.32 \pm 0.03$	      \\[1.5mm]
$\chi ^2$/d.o.f.		   	   & 47.59/40						 & 38.07/35			& 56.08/40	          \\[1.5mm]
Flux ($\ergcms$)               & $1.500^{+0.006}_{-0.01} \E{-9}$ & $3.97 \pm 0.01 \E{-10}$ 	& $6.51 \pm 0.08 \E{-11}$        \\
\hline
\end{tabular}
\end{table*}

For all ObsIDs viz. 70131-03-01-00 (HF), 70131-01-09-000 (LF) and 70131-02-04-00 (BE), data was extracted
from PCU2. For the HF and BE, the spectra were fit with 
{\it tbabs(powerlaw+bbody)}. The LF spectrum was fit with a simple absorbed powerlaw. 
Additionally, a narrow Gaussian component centred at $\sim 6.5$~keV was required 
for the low flux observations. The neutral hydrogen column density was fixed to 
$N_H = 1\times10^{22} \mbox{ cm}^{-2}$ \citep{mille03,gierl05}.  The results are 
presented in Table~\ref{tab.J1751}.

\subsection{SAX J1748.9--2021:}

\begin{table*}
\centering
\caption[%
    Spectral fit parameters of SAX~J1748.9--2021.
]{\small%
    Spectral fit parameters of SAX~J1748.9--2021, see Table~\ref{tab.J1756}
    for details.
}
\begin{tabular}{|l  | l  l  l |}
\hline
 & \multicolumn{3}{c|}{\bf SAX~J1748.9--2021} \\
 & High flux & Low flux & Background \\
\hline 
ObsIDs                         & 94315-01-06-07				& 60035-02-03-02	 		& 94315-01-11-02    \\[1.5mm]
$N_H (10^{22}\mbox{ cm}^{-2})$ & 0.59 (fixed)               & 0.59 (fixed)       			& 0.59 (fixed)		\\[1.5mm]
$T_{\rm bb}$	(keV)          & $2.11 \pm 0.14$			& $0.63^{+0.06}_{-0.11}$		& --			\\[1.5mm]
${\rm Norm}_{\rm bb}$          & $1.30 \pm 0.19 \E{-2}$	    & $6.5^{+2.9}_{-1.7} \E{-3}$	& --			\\[1.5mm]
$\Gamma$ (cutoffpl)            & $1.00 \pm 0.16$			& $0.39^{+0.15}_{-0.23}$		& --			\\[1.5mm]
$E_{\rm cut}$ (keV)            & $4.1  \pm 0.6$     		& $3.36^{+0.16}_{-0.21}$		& --			\\[1.5mm]
$\Gamma$ (power law)           & --							& --							& $2.2 \pm 0.2$    \\[1.5mm]
$ \chi ^2$/d.o.f.              & 45.41/41					& 36.21/33						& 28.23/43			\\[1.5mm]
Flux ($\ergcms$)               & $4.13 \pm 0.01 \E{-9}$     & $2.960 \pm 0.005 \E{-9}$   	& $1.9 \pm 0.2 \E{-11}$    \\
\hline
\end{tabular}
\label{tab.J1748}
\end{table*}

For ObsID 94315-01-06-07 (HF) and  60035-02-03-02 (LF) the data were extracted 
from PCU 2, whereas for 94315-01-11-02 (BE) the combined spectra from
PCU 0 and 2 were used. The HF and LF spectra fit with {\it tbabs(bbody+cutoffpl)}. 
The BE spectrum was fit using an absorbed power-law. The hydrogen column density 
was fixed to $N_H = 5.9 \times 10^{21} \mbox{ cm}^{-2}$ \citep{harri96}. A 
Gaussian emission feature centred at $\sim 6.5$~keV was used to improve 
statistics of the LF spectrum fit. The results are presented in
Table~\ref{tab.J1748}.

\subsection{Swift J1749.4--2807:}

\begin{table*}
\centering
\caption[%
    Spectral fit parameters of Swift~J1749.4--2807.
]{\small%
    Spectral fit parameters of Swift~J1749.4--2807, see Table~\ref{tab.J1756}
    for details.
}
\label{tab.J1749}
\begin{tabular}{|l  | l  l  l |}
\hline
 & \multicolumn{3}{c|}{\bf Swift~J1749.4--2807} \\
 & High flux & Low flux & Background \\
\hline 
ObsIDs                         & 95085-09-01-00          	& 95085-09-02-05          	& 95085-09-02-10       \\[1.5mm]
$N_H (10^{22}\mbox{ cm}^{-2})$ & 3 (fixed)               	& 3 (fixed)               	& 3 (fixed)            \\[1.5mm]
$T_{\rm bb}$~(keV)             & --                      	& $1.67 \pm 0.11$        	& $1.8 \pm 0.2$	      \\[1.5mm]
${\rm Norm}_{\rm bb}$          & --	                     	& $7.94 \pm 1.2 \E{-4}$   	& $5.8 \pm 1.2 \E{-4}$ \\[1.5mm]
$\Gamma$                       & $1.89 \pm 0.02$         	& $1.78 \pm 0.08$        	& $1.90 \pm 0.09$      \\[1.5mm]
$ \chi ^2$/d.o.f.              & 43.21/41                	& 54.01/33	               	& 39.77/39             \\[1.5mm]
Flux ($\ergcms$)               & $5.24 \pm 0.06 \E{-10}$ 	& $2.67 \pm 0.02 \E{-10}$ 	& $2.41 \pm 0.03 \E{-10}$        \\
\hline
\end{tabular}
\end{table*} 

The data were extracted from PCU2 for all ObsIDs (HF: 95085-09-01-00, LF:
95085-09-02-07, BE: 95085-09-02-10). There was an eclipse during the
observation in 95085-09-01-00 \citep{markw10a,ferri11}, so for the spectral
analysis of that observation we considered only the data from the initial 990~s
when the source was visible. The spectra were fit with power-law continuum model. 
The neutral hydrogen column density was fixed to $N_H = 3 \E{22} \mbox{ cm}^{-2}$ 
\citep{ferri11,wijna09}. For the LF and BE an additional blackbody component was required. 
The results are presented in Table~\ref{tab.J1749}.

\subsection{Aql X-1:}

\begin{table*}
\centering
\caption[%
    Spectral fit parameters of Aql X-1.
]{\small%
    Spectral fit parameters of Aql X-1, see Table~\ref{tab.J1756}
    for details.
}
\label{tab.aql}
\begin{tabular}{|l  | l  l |}
\hline
 & \multicolumn{2}{c|}{\bf Aql X-1} \\
 & High/Low flux &  Background \\
\hline 
ObsIDs                         & 30188-03-05-00          & 30073-06-01-00  \\[1.5mm]
$N_H (10^{22}\mbox{ cm}^{-2})$ & 0.4 (fixed)             & 0.4 (fixed)  \\[1.5mm]
$T_{\rm bb}$~(keV)             & $0.81^{+0.44}_{-0.09}$  & --  \\[1.5mm]
${\rm Norm}_{\rm bb}$          & $1.2^{+2.1}_{-0.6}$	 & --  \\[1.5mm]
$\Gamma$		       		   & --			 			 & $1.92 \pm 0.02$  \\[1.5mm]
$\Gamma$ (cutoffpl)            & $0.7^{+0.2}_{-0.4}$  	 & --  \\[1.5mm]
$E_{\rm cut}$ (keV)            & $3.4 \pm 0.4$         	 & --  \\[1.5mm]
$E_{1}$       (keV)            & $6.32^{+0.18}_{-0.38}$  & --  \\[1.5mm]
$\sigma_{1}$  (keV)            & $1.0 \pm 0.3$           & --  \\[1.5mm]
$ \chi ^2$/d.o.f.              & 45.95/38                & 45.39/43  \\[1.5mm]
Flux ($\ergcms$)               & $8.74 \pm 0.01 \E{-9}$  & $1.34 \pm 0.05 \E{-11}$  \\
\hline
\end{tabular}
\end{table*} 

Since the pulsations of Aql~X-1 have been detected only once (ObsID 30188-03-05-00),
we use this observation as both the HF and LF. We extract the
data from PCU2 and fit the spectrum with a {\it tbabs(cutoffpl+Gaussian+bbody)}
model. The background emission is evaluated from ObsID 30073-06-01-00, which was 
fit with a simple absorbed power-law. Our results are shown in Table~\ref{tab.aql}.

\subsection{IGR~J00291+5934:}

\begin{table*}
\centering
\caption[%
    Spectral fit parameters of IGR~J00291+5934, see Table~\ref{tab.J1756}
]{\small%
    Fit parameters from spectral analysis of IGR~J00291+5934.
    }
\label{tab.J00291}
\begin{tabular}{|l  | l  l  l |}
\hline
 & \multicolumn{3}{c|}{\bf IGR~J00291+5934}     \\
 & High flux & Low flux & Background \\
\hline 
ObsIDs                         & 90052-03-01-00          & 90425-01-02-01          	& 90425-01-03-06      \\[1.5mm]
$N_H (10^{22}~\mbox{cm}^{-2})$ & 0.46 (fixed)            & 0.46 (fixed)            	& 0.46 (fixed)   	 \\[1.5mm]
$\Gamma$                       & $1.50 \pm 0.04$         & $1.73 \pm 0.04$         	& $1.50 \pm 0.03$    \\[1.5mm]
$T_{\rm BB}$~(keV)             & $1.14 \pm 0.06$         & --			   			& --  			     \\[1.5mm]
${\rm Norm}_{\rm bb}$          & $1.05 \pm 0.03 \E{-3}$  & --			   			& --	     		     \\[1.5mm]
$\chi^2$/d.o.f.                & 48.81/42                & 22.92/35		   			& 30.55/41		     \\[1.5mm]
Flux ($\ergcms$)               & $9.7 \pm 0.05 \E{-10}$ & $1.09 \pm 0.02 \E{-10}$ 	& $5.76 \pm 0.09 \E{-11}$       \\
\hline 
\end{tabular} 
\end{table*} 

For the ObsID 90052-03-01-00 (HF) data from PCU2 were used
for spectral analysis. For 90425-01-02-01 (LF) and
90425-01-03-06 (BE) data from all active PCUs (PCU
0, 2 and PCU 0, 2, 3 respectively) were combined to improve the photon
statistics. 

The  HF spectrum was fit with the thermal blackbody and power-law models. The LF and BE spectra were fit with 
a simple absorbed power law. The neutral hydrogen column density was
fixed to $N_H = 0.46 \times 10^{22} \mbox{ cm}^{-2}$ based on measurements
from {\it XMM-Newton} and {\it Chandra} observations \citep{torre08,paizi05}. A gaussian emission feature was required to improve the fits of the LF and BE spectra.
The fit results are given in Table~\ref{tab.J00291}. 

\clearpage

\end{document}